\documentclass[twocolumn,iop]{emulateapj}

\usepackage{graphics,color}
\graphicspath{{Figures/}{./}}

\newcommand{\code}[1]{\texttt{#1}}
\newcommand{\Ks}{{\ensuremath K_s}}

\begin{document}

\title{The Tenth Data Release of the Sloan Digital Sky Survey:
First Spectroscopic Data from the \mbox{SDSS-III} Apache Point
Observatory Galactic Evolution Experiment}

\author{
Christopher~P.~Ahn\altaffilmark{1},    
Rachael~Alexandroff\altaffilmark{2},
Carlos~{Allende~Prieto}\altaffilmark{3,4},
Friedrich~Anders\altaffilmark{5,6},
Scott~F.~Anderson\altaffilmark{7},
Timothy~Anderton\altaffilmark{1},
Brett~H.~Andrews\altaffilmark{8},
\'Eric~Aubourg\altaffilmark{9},
Stephen~Bailey\altaffilmark{10},
Fabienne~A.~Bastien\altaffilmark{11}
% ...
}

\altaffiltext{1}{
Department of Physics and Astronomy, 
University of Utah, Salt Lake City, UT 84112, USA.
}

\altaffiltext{2}{
Center for Astrophysical Sciences, 
Department of Physics and Astronomy, 
Johns Hopkins University, 3400 North Charles Street, Baltimore, MD 21218, USA.
}

\altaffiltext{3}{
Instituto de Astrof{\'\i}sica de Canarias (IAC), C/V{\'\i}a L\'actea,
s/n, E-38200, La Laguna, Tenerife, Spain.
}

\altaffiltext{4}{
Departamento de Astrof\'{\i}sica, 
Universidad de La Laguna, 
E-38206, La Laguna, Tenerife, Spain.
}

\altaffiltext{5}{
Leibniz-Institut f\"ur Astrophysik Potsdam (AIP), An der Sternwarte 16, 
D-14482 Potsdam, Germany.
}

\altaffiltext{6}{
Technische Universit\"at Dresden (TUD), 
Institut f\"ur Kern- und Teilchenphysik, 
D-01062 Dresden, Germany.
}

\altaffiltext{7}{
Department of Astronomy, University of Washington, 
Box 351580, Seattle, WA 98195, USA.
}

\altaffiltext{8}{
Department of Astronomy, 
Ohio State University, 140 West 18th Avenue, Columbus, OH 43210, USA.
}

\altaffiltext{9}{
APC, University of Paris Diderot, CNRS/IN2P3, CEA/IRFU, Observatoire de Paris, Sorbonne Paris Cit\'e, F-75205 Paris, France.
}

\altaffiltext{10}{
Lawrence Berkeley National Laboratory, One Cyclotron Road,
Berkeley, CA 94720, USA.
}

\altaffiltext{11}{
Department of Physics and Astronomy, Vanderbilt University, 
VU Station 1807, Nashville, TN 37235, USA.
}

% ...

\shorttitle{SDSS DR10}

\begin{abstract}

The Sloan Digital Sky Survey (SDSS) has been in operation since 2000
April.  This paper presents the tenth public data release (DR10) from its
current incarnation, \mbox{SDSS-III}.  This data release includes
the first spectroscopic data from the Apache Point Observatory Galaxy
Evolution Experiment (APOGEE), along with spectroscopic
data from the Baryon Oscillation Spectroscopic Survey (BOSS) taken
through 2012 July.
The APOGEE instrument is a near-infrared $R\sim$~22,500 300-fiber
spectrograph covering $1.514$--$1.696$~$\mu$m.  The APOGEE survey is studying
the chemical abundances and radial velocities of roughly 100,000 red giant
star candidates in the bulge, bar, disk, and halo of the Milky Way. DR10
includes 178,397 spectra of 
57,454 stars, each typically observed three or more times, from
APOGEE. Derived quantities from these spectra (radial velocities,
effective temperatures, surface gravities, and metallicities) are also
included.

DR10 also roughly doubles the number of BOSS spectra over those
included in the ninth data release.  DR10 includes
a total of 1,507,954 BOSS spectra, comprising 927,844 galaxy spectra;
182,009 quasar spectra; and 159,327 stellar spectra, selected over
6373.2~deg$^2$.

\end{abstract}

\section{Introduction}
\label{sec:introduction}

The Sloan Digital Sky Survey (SDSS) has been in continuous operation
since 2000 April.  It uses a dedicated wide-field 2.5-m telescope
\citep{Gunn06} at Apache Point Observatory (APO) in the Sacramento
Mountains in Southern New Mexico.  It was originally instrumented with
a wide-field imaging camera with an effective area of 1.5~deg$^2$
\citep{Gunn98}, and a pair of double spectrographs fed by 640 fibers
\citep{Smee13}.  The initial survey \citep{York00} carried out imaging
in five broad bands ($ugriz$) \citep{Fukugita96} to a depth of $r\sim 22.5$~mag over 11,663~deg$^2$ of high-latitude sky, and spectroscopy of 1.6 million galaxy,
quasar, and stellar targets over 9380~deg$^2$.  The resulting images
were calibrated astrometrically \citep{Pier03} and
photometrically \citep{Ivezic04,Tucker06,Padmanabhan08}, and the
properties of the detected objects were measured \citep{Lupton01}.
The spectra were calibrated and redshifts and classifications
determined \citep{Bolton12}.  The data have been released publicly in a
series of roughly annual data releases (\citealt{EDR, DR1, DR2, DR3,
  DR4, DR5, DR6, DR7}; hereafter EDR, DR1, DR2, DR3, DR4, DR5, DR6,
DR7, respectively) as the project went through two funding phases, termed SDSS-I
(2000--2005) and SDSS-II (2005--2008).

In 2008, the SDSS entered a new phase, designated \mbox{SDSS-III}
\citep{Eisenstein11}, in which it is currently operating.  SDSS-III has four
components.  The Sloan Extension for Galactic Understanding and
Exploration 2 (SEGUE-2), an expansion of a similar project carried out
in SDSS-II \citep{Yanny09}, used the SDSS spectrographs to obtain
spectra of about 119,000 stars, mostly at high Galactic latitudes.  The Baryon
Oscillation Spectroscopic Survey (BOSS; \citealt{Dawson13}) rebuilt the spectrographs to
improve throughput and increase the number of fibers to 1000 \citep{Smee13}.
BOSS enlarged the imaging footprint of SDSS to 14,555~deg$^2$, and is
obtaining spectra of galaxies and quasars with the primary goal
of measuring 
the oscillation signature in the clustering of matter as a cosmic
yardstick to constrain cosmological models.  The Multi-Object APO
Radial Velocity Exoplanet Large-area Survey (MARVELS), which finished
its data-taking in 2012, used a 60-fiber
interferometric spectrograph to measure high-precision radial
velocities of stars in a search for planets and brown dwarfs.
Finally, the Apache Point Observatory Galactic Evolution Experiment
(APOGEE) uses a 300-fiber spectrograph to observe bright ($H<13.8$~mag)
stars in the $H$ band at high resolution ($R\sim $~22,500) for accurate
radial velocities and detailed elemental abundance determinations.

We have previously had two public data releases of data from \mbox{SDSS-III}.
The Eighth Data Release (DR8; \citealt{DR8}) included all data from
the SEGUE-2 survey, as well as $\sim 2500$~deg$^2$ of new imaging
data in the Southern Galactic Cap as part of BOSS.  The Ninth Data
Release (DR9, \citealt{DR9}) included the first spectroscopic data from the
BOSS survey: over 800,000 spectra selected from 3275~deg$^2$ of sky.

  This paper describes the Tenth Data Release (hereafter DR10) of the
  SDSS survey.  This release includes almost 680,000 new BOSS spectra, covering
  an additional 3100~deg$^2$ of sky.  It also includes the first public release
  of APOGEE spectra, with almost 180,000 spectra of more than 57,000 
  stars in a wide range of Galactic environments.  As in previous SDSS
  data releases, DR10 is cumulative; it includes all data that were
  part of DR1--9.  All data released with DR10 are publicly available
  on the \mbox{SDSS-III} website\footnote{\url{http://www.sdss3.org/dr10/}}
  and links from it.  

  The scope of the data release is described in detail in
  Section~\ref{sec:scope}.  We describe the APOGEE data in
  Section~\ref{sec:apogee}, and the new BOSS data in Section~\ref{sec:boss}.
  The mechanisms for data access are described in
  Section~\ref{sec:distribution}.  We outline the future of SDSS in Section~\ref{sec:future}.

\section{Scope of DR10}
\label{sec:scope}

DR10 presents the release of the first year of data from the
\mbox{SDSS-III} APOGEE infrared spectroscopic survey and the first 2.5
years of data from the \mbox{SDSS-III} BOSS optical spectroscopic
survey.  In each case these data extend to the 2012 telescope shutdown
for the summer monsoon season.  

APOGEE was commissioned from 2011 May up through the summer shutdown
in 2011 July.  Survey-quality observations began 2011 Aug 31 (UTC-7),
corresponding to Modified Julian Date (MJD) 55804.
The APOGEE data presented in DR10 include all commissioning and
survey data taken up to and including MJD 56121 
(2012 July 13).  However, detailed stellar parameters are only
presented for APOGEE spectra obtained after commissioning was complete.  
The BOSS data include all data
taken up to and including MJD 56107
(2012 June 29).

DR10 also includes the imaging and spectroscopic data from SDSS-I/II
and \mbox{SDSS-III} SEGUE-2, the imaging data for the BOSS Southern
Galactic Cap first presented in DR8, as well as the spectroscopy from
the first 2.5 years of BOSS.  Table~\ref{table:dr10_contents} lists
the contents of the data release, including the imaging coverage and
number of APOGEE and BOSS plates and spectra.  APOGEE plates are
observed multiple times (``visits'') to build signal-to-noise ratio
(S/N) and to search for radial velocity variations; thus the number of
spectra in DR10 is significantly larger than the number of unique stars
observed.  While there are fewer repeat spectra in BOSS, we sill 
distinguish between the total number of spectra, and the number of
unique objects observed in BOSS as well.  The numbers for the imaging data, unchanged since DR8, 
also distinguish between unique and total area and number of detected
objects.  The multiple repeat observations of the Equatorial Stripe in
the Fall sky \citep{Annis11}, used to search for Type Ia supernovae
\citep{Frieman08}, dominate the difference between total and unique
area imaged.

New in DR10 are morphological classifications of SDSS images of
galaxies by 200,000 citizen scientists via the Galaxy Zoo project
\citep{Lintott08,Lintott11,Willett13}.  These classifications include
both the basic (spiral--early-type) morphologies for all
$\sim$1~million galaxies from the SDSS-I/II Main Galaxy Sample
\citep{Strauss02}, as well as more detailed classifications of the
internal structures in the brightest 250,000 galaxies.

\begin{figure*}
\plotone{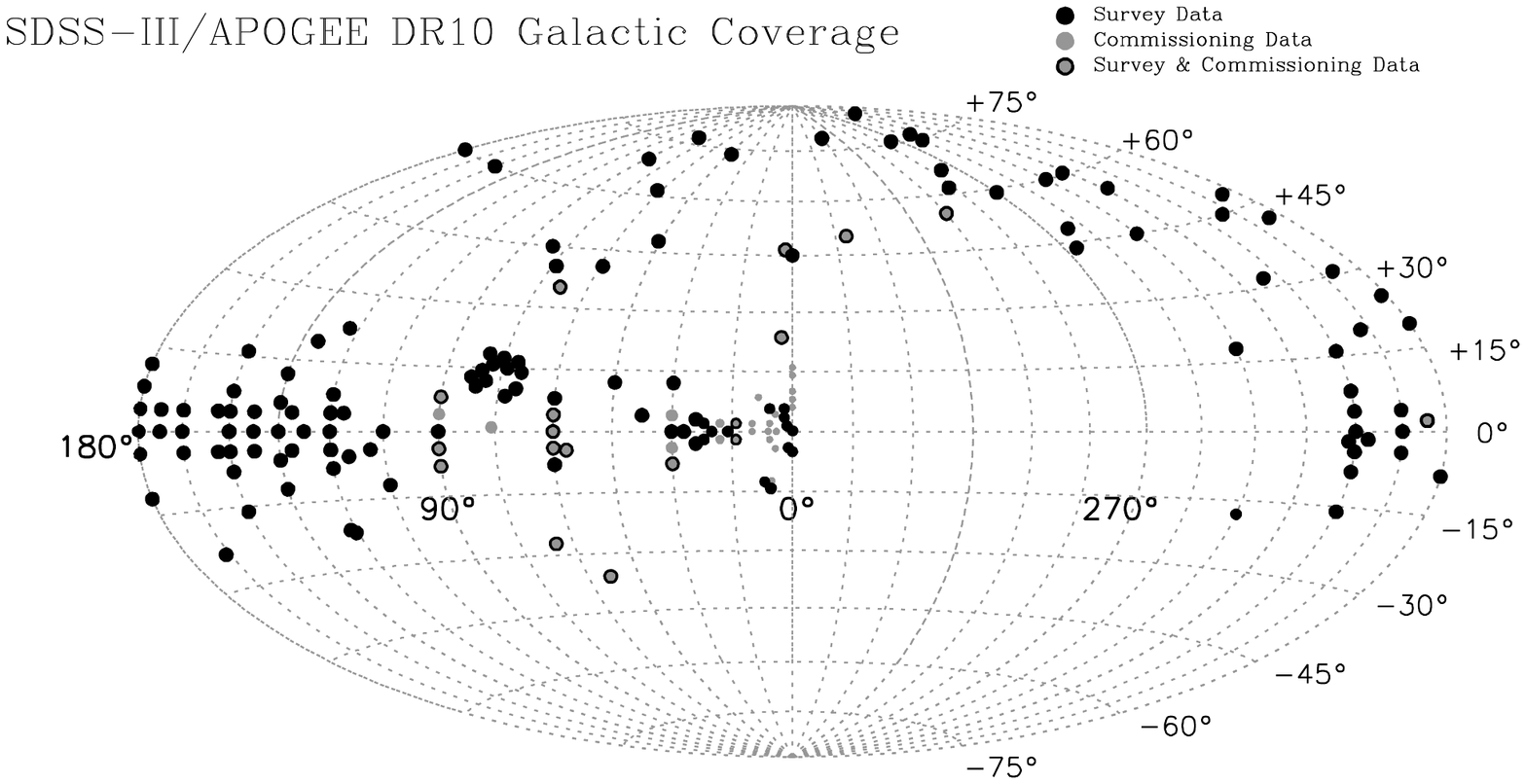}
\plotone{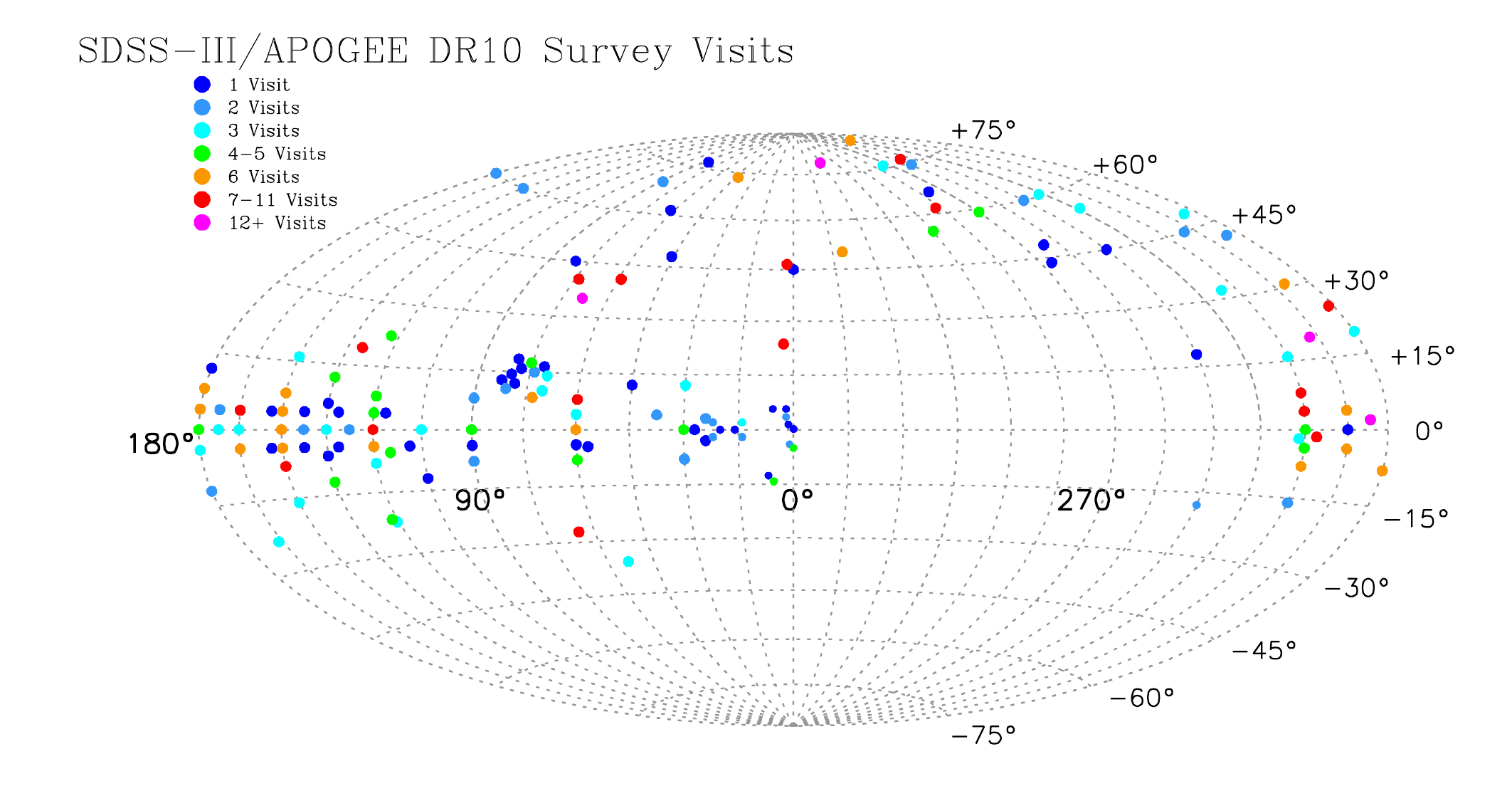}
\caption{The distribution on the sky of all APOGEE DR10 pointings 
  in Galactic
  coordinates: the Galactic Center is in the middle of the diagram.  Each circle represents a
  pointing.  APOGEE often has several distinct plates for a single location on the sky; DR10 includes 
  170 locations, which are shown above.  Smaller circles (primarily
  near the Galactic Center) represent
  locations where plates were drilled over only a fraction of the 7~deg$^2$ focal plane to
  minimize differential atmospheric refraction.  Note the
  concentration of fields along the Galactic Plane.  The concentration of
  pointings at
$l= 75^\circ, b= +15^\circ$ is a special program targeting stars
  observed by the {\it Kepler} telescope; see Section~\ref{sec:apogee_apokasc}.  
(top) Distribution of pointings in both the commissioning and
  survey phases (both are included in DR10). 
(bottom) Pointings distinguished by the number of visits obtained by DR10 in the survey phase.
}
\label{fig:apogee_skydist}
\end{figure*}

\begin{figure*}
\plotone{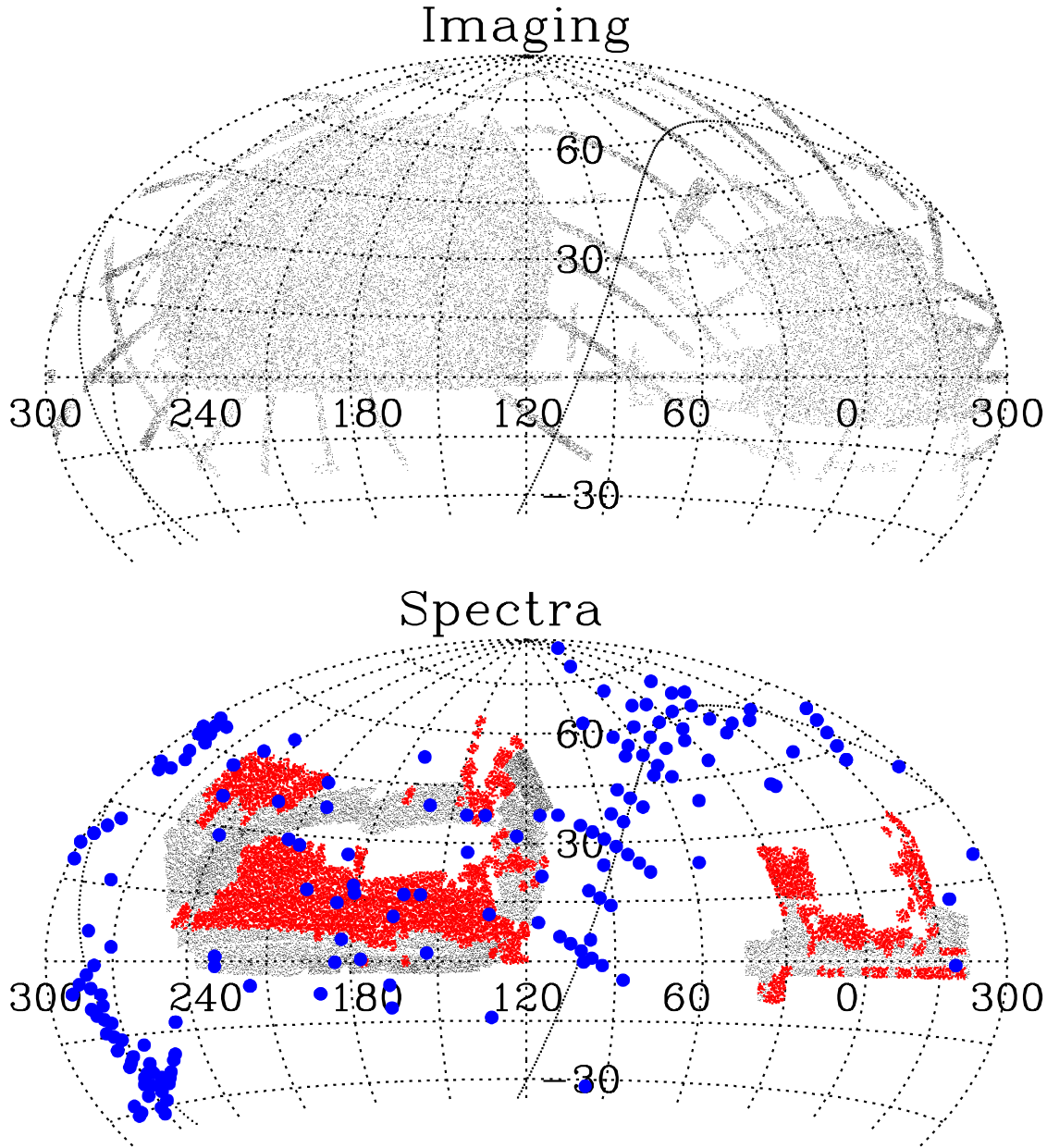}
\caption{The distribution on the sky of all SDSS imaging (top;
  14,555~deg$^2$ -- same as DR8 and DR9) and BOSS and APOGEE DR10
  spectroscopy (bottom; 6373.2~deg$^2$) in J2000 equatorial coordinates
  ($\alpha=0^\circ$ is right of center in this projection).  Grey
  shows regions included in DR9; the increment included in DR10 is in
  red.  The blue shows the positions of APOGEE pointings included in
  DR10.  The Galactic Plane is shown by the dotted line.  The Northern
  Galactic Cap is on the left of the figure, and the Southern Galactic
  Cap on the right.  The BOSS sky coverage shown is actually constructed using a random
  subsample of the BOSS DR10Q quasar catalog \citep{Paris13}.  
  The sky below $\delta<-30^\circ$ is never at an airmass of
  of less than 2.0 from APO (latitude=+32$^\circ$46\arcmin49\arcsec).
}
\label{fig:skydist_boss}
\end{figure*}

The celestial footprint of the APOGEE spectroscopic coverage in DR10
is shown in Figure~\ref{fig:apogee_skydist} in Galactic coordinates;
Figure~\ref{fig:skydist_boss} repeats this in Equatorial coordinates,
and shows the imaging and BOSS spectroscopy sky coverage as well.  The
distribution on the sky of SDSS-I/II and SEGUE-2 spectroscopy 
is not shown here; see the DR7 and DR8 papers.  
APOGEE fields span all of the Galactic components visible
from APO, including the Galactic center and disk, as well as
fields at high Galactic latitudes
to probe the halo.  The Galactic center observations occur at high
airmass, thus the differential atmospheric refraction
across the field of view changes rapidly with hour angle.  
Therefore targets in these fields are not distributed over the full 7~deg$^2$ of each plate, but rather over a
smaller region from 0.8 to 3.1~deg$^2$, as indicated by the smaller dots
in Figure~\ref{fig:apogee_skydist}.  The clump of points centered
roughly at $l= 75^\circ, b= +15^\circ$ are special plates targeting
stars previously observed by NASA's {\it Kepler} mission, as described in detail in
Section~\ref{sec:apogee_apokasc}.

The additional BOSS spectroscopy fills in most of the ``doughnut''
defined by the DR9 coverage in the North Galactic Cap.  The DR10 BOSS sky
coverage relative to the 10,000~deg$^2$ full survey region is
described further in Section~\ref{sec:boss}. 

\begin{deluxetable}{lrrr}
\tablecolumns{4}
\tablecaption{Contents of DR10\label{table:dr10_contents}}
\startdata
\cutinhead{Optical Imaging\tablenotemark{a}}
& & Total & Unique\tablenotemark{b} \\
\multicolumn{2}{l}{Area Imaged [deg$^2$]}  & 31637 &     14555 \\  \multicolumn{2}{l}{Cataloged Objects} & 1231051050 & 469053874 \\   \cutinhead{APOGEE Spectroscopy}
                          & Commiss. & Survey & Total \\
Plate-Visits              &            98 &    586 &   684 \\ Plates                    &            51 &    232 &   281 \\ Pointings                 &            43 &    150 &   170 \\
& & & \\
& & Spectra & Stars \\
                                                 \multicolumn{2}{l}{All Stars\tablenotemark{c}}             & 178397 & 57454 \\ \multicolumn{2}{l}{Commissioning Stars}      &  24943 & 11987 \\ 
\multicolumn{2}{l}{Survey Stars}             & 153454 & 47452 \\ 
\multicolumn{2}{l}{\quad Stars with S/N $> 100$\tablenotemark{d}} &  \nodata & 47675 \\
\multicolumn{2}{l}{\quad Stars with $\ge3$ visits}      &  \nodata & 29701 \\
\multicolumn{2}{l}{\quad Stars with $\ge12$ visits}     &  \nodata &  923 \\
\multicolumn{2}{l}{\quad Stellar parameter standards} &  5178 & 1065 \\
\multicolumn{2}{l}{\quad Radial velocity standards}   &   162 &   16 \\
\multicolumn{2}{l}{\quad Telluric line standards}     & 24283 & 7003 \\
\multicolumn{2}{l}{\quad Ancillary science program objects}  &  8894 & 3344 \\

\cutinhead{BOSS Spectroscopy}

& & Total & Unique\tablenotemark{b}  \\
\multicolumn{2}{l}{Spectroscopic effective area [deg$^2$]}              & \nodata & 6373.2  \\
\multicolumn{2}{l}{Plates\tablenotemark{e}}                   &  1515    &  1489     \\
\multicolumn{2}{l}{Optical Spectra observed\tablenotemark{f}} & 1507954 & 1391792  \\
& & &  \\
\multicolumn{2}{l}{All Galaxies}                          & 927844 & 859322  \\
\multicolumn{2}{l}{\quad CMASS\tablenotemark{g}       }   & 612195 & 565631  \\
\multicolumn{2}{l}{\quad LOWZ\tablenotemark{g}        }   & 224172 & 208933  \\
\multicolumn{2}{l}{All Quasars}                           & 182009 & 166300  \\
\multicolumn{2}{l}{\quad Main\tablenotemark{h}       }    & 159808 & 147242  \\
\multicolumn{2}{l}{\quad Main, $2.15<z<3.5$\tablenotemark{i} } & 114977 & 105489  \\
\multicolumn{2}{l}{Ancillary program spectra}             &  72184 &  65494  \\
\multicolumn{2}{l}{Stars}                                 & 159327 & 144968  \\
\multicolumn{2}{l}{\quad Standard stars}                  &  30514 &  27003  \\
\multicolumn{2}{l}{Sky spectra}                           & 144503 & 138491  \\
\multicolumn{2}{l}{Unclassified spectra\tablenotemark{j}} & 101550 &  89003  \\
\cutinhead{All Optical Spectroscopy from SDSS up through DR10}
\multicolumn{2}{l}{Total spectra} & 3358200  \\ \multicolumn{2}{l}{Total useful spectra\tablenotemark{k}} & 3276914  \\ 
\multicolumn{2}{l}{ \quad  Galaxies}   & 1848851  &  \\ \multicolumn{2}{l}{ \quad  Quasars}    &  316125  &  \\ \multicolumn{2}{l}{ \quad  Stars}      &  736484  &  \\ \multicolumn{2}{l}{ \quad  Sky}        &  247549  &  \\ \multicolumn{2}{l}{ \quad  Unclassified\tablenotemark{j} } &  138663 \\ \enddata
\tablenotetext{a}{These numbers are unchanged since DR8.}
\tablenotetext{b}{Removing all duplicates, overlaps, and repeat visits from the ``Total'' column.}
\tablenotetext{c}{2,155 stars were observed both during the commissioning and survey phases.  The co-added spectra are kept separate between these two phases.  Thus the number of coadded spectra is greater than the number of unique stars observed.}
\tablenotetext{d}{Signal-to-noise ratio per half resolution element $>100$.}
\tablenotetext{e}{Twenty-six plates of the 1515 observed plates were
  re-plugged and re-observed for calibration purposes. Six of the 1489
  unique plates are different drillings of the same set of objects.}
\tablenotetext{f}{This excludes the small fraction of the observations
  through fibers that are broken or that fell out of their holes after
  plugging.  There were 1,515,000 spectra attempted.}
\tablenotetext{g}{``CMASS'' and ``LOWZ'' refer to the two galaxy
  target categories used in BOSS \citep{DR9}.  They are both
  color-selected, with LOWZ galaxies in the redshift range $0.15 < z <
  0.4$, and CMASS galaxies in the range $0.4 < z
  < 0.8$.}
\tablenotetext{h}{This counts only quasars that were targeted by the
  main quasar survey \citep{Ross12},
  and thus does not include those from ancillary programs
  \citep{Dawson13}.} \tablenotetext{i}{Quasars with redshifts in the range $2.15<z<3.5$
  provide the most signal in the BOSS spectra of the
  Ly-$\alpha$ forest.}
\tablenotetext{j}{Non-sky spectra for which the automated
redshift/classification pipeline \citep{Bolton12} gave no reliable
classification, as indicated by the \code{ZWARNING} flag.}
\tablenotetext{k}{Spectra on good or marginal plates.}
\end{deluxetable}

\section{The Apache Point Observatory Galaxy Evolution Experiment (APOGEE)}
\label{sec:apogee}

\subsection{Overview of APOGEE}
\label{sec:apogee_overview}

Stellar spectra of red giants in the $H$ band (1.5--1.8~$\mu$m) show a
rich range of absorption lines from a wide variety of elements.  At
these wavelengths, the absorption due to dust in the plane of the
Milky Way is much reduced compared to that in the optical bands.  A
high-resolution study of stars in the $H$ band allows studies of all
components of the Milky Way, across the disk, in the bulge, and out to the halo.

APOGEE's goal is to trace the history of star formation in, and the
assembly of, the Milky Way by obtaining $H$-band spectra of 100,000 red
giant candidate stars throughout the Galaxy.  Using an infrared multi-object spectrograph
with a resolution of $R\equiv\lambda/\Delta\lambda \sim$~22,500, APOGEE can survey the halo, disk,
and bulge in a much more uniform fashion than previous surveys.
The APOGEE spectrograph features a 50.8~cm $\times$ 30.5~cm mosaiced
volume-phase holographic (VPH) grating and a six-element camera having
lenses with a
maximum diameter of 40~cm. 
APOGEE takes advantage
  of the fiber infrastructure on the SDSS telescope, using 300 fibers, each
  subtending $2$\arcsec\ on the sky, distributed over the full 7~deg$^2$
  field of view (with the exception of plates observed at high
  airmass, as noted above).  The spectrograph itself sits in a
  temperature-controlled room, and thus does not move with the
  telescope. 
  The light from the fibers falls onto three
  \mbox{HAWAII-2RG} \mbox{$\rm 2K \times 2K$} infrared detectors \citep{Garnett04,Rieke07},
  that cover the wavelength range from 1.514~$\mu$m to 1.696~$\mu$m, with
  two gaps (see Section~\ref{sec:apogee_instrument} for details).
  APOGEE targets are chosen with magnitude and color cuts from
  photometry of the Two-Micron All-Sky Survey (2MASS;
  \citealt{Skrutskie06}), with a median $H=10.9$~mag and with 99.6\% of the stars brighter than $H=13.8$~mag (on the 2MASS Vega-based system).

  The high resolution of the spectra and the stability of the instrument allow accurate radial velocities
  with a typical uncertainty of 100~m~s$^{-1}$,
  and
  detailed abundance determinations for approximately 15
  chemical elements. 
  In addition to being key in identifying binary star systems, the
  radial velocity data are being used to explore the kinematical 
  structure of the Milky Way and its substructures \citep[e.g., ][]{Nidever12} and to constrain
  dynamical models of its disk \citep[e.g., ][]{Bovy12}. The chemical abundance data allow studies of
  the chemical evolution of the Galaxy \citep{GarciaPerez13a} and the history of star
  formation. The combination of kinematical and chemical data will allow
  important new constraints on the formation history of the Milky
  Way.

A full overview of the APOGEE survey will be presented in
S.~Majewski et al. (2014, in preparation).
  The APOGEE instrument will be detailed in
J.~Wilson et al. (2014, in preparation) and is summarized here in
Section~\ref{sec:apogee_instrument}.  The target selection process for
APOGEE is described in \citet{Zasowski13} and is presented in brief
here in Section~\ref{sec:apogee_targeting}.  In
Section~\ref{sec:apogee_apokasc} we describe a unique
cross-targeting program between \mbox{SDSS-III} APOGEE and
asteroseismology measurements from the NASA {\it Kepler}
telescope\footnote{\url{http://kepler.nasa.gov/}} \citep{Gilliland10}.
Section~\ref{sec:apogee_analysis} describes the reduction pipeline
that processes the APOGEE data
and produces calibrated one-dimensional spectra of each star, including
accurate radial velocities (D.~Nidever et al., 2014, in preparation). Important caveats regarding APOGEE data of which potential
users should be aware are described in
Section~\ref{sec:apogee_caveats}.  Section~\ref{sec:apogee_aspcap}
describes the pipeline that measures stellar properties and elemental
abundances -- the APOGEE Stellar Parameters and Chemical Abundances
Pipeline (ASPCAP;
 M.~Shetrone et al., 2014, in preparation;
 A.~Garc\'\i{}a-P\'erez et al., 2014, in preparation,
 \citealt{Meszaros13}).
Section~\ref{sec:apogee_data} summarizes the APOGEE data products
available in DR10.

\begin{figure*}
\plotone{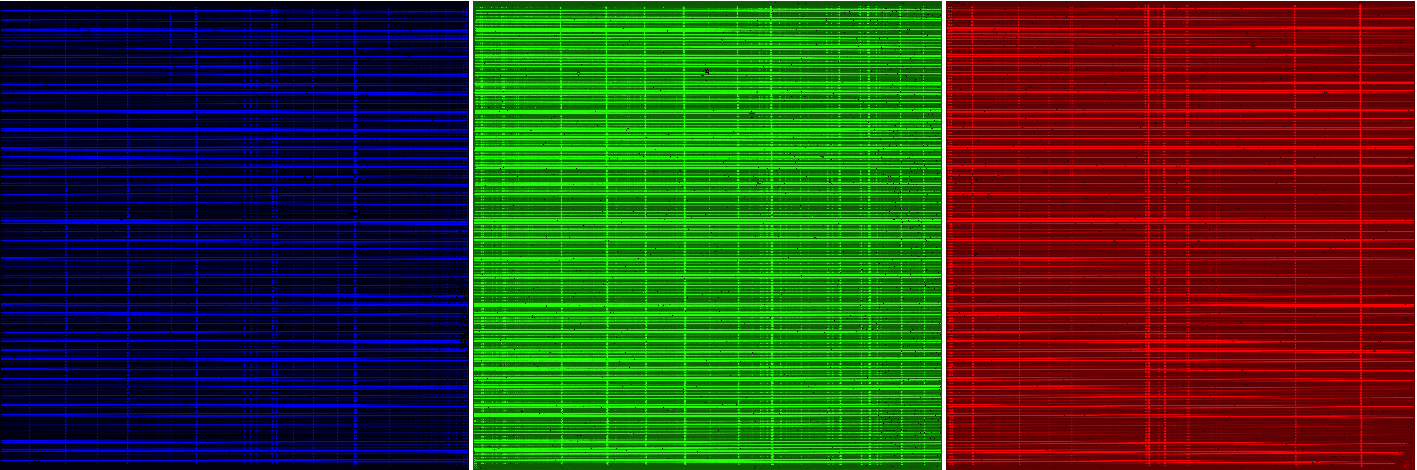}
\plotone{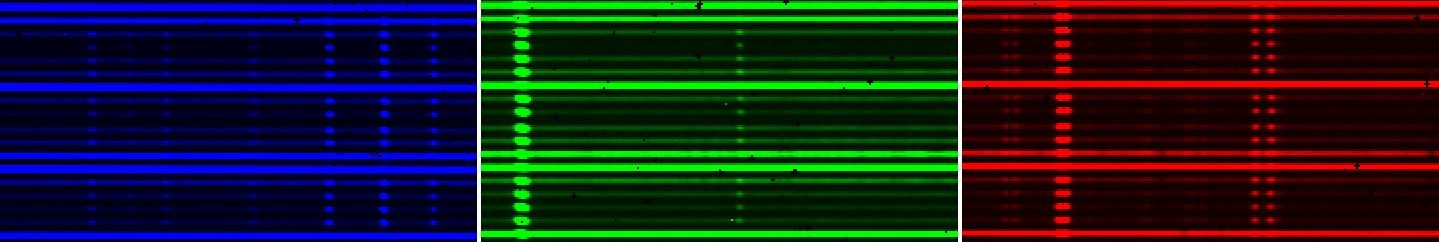}
\caption{(top) A 2D spectrogram from the APOGEE instrument.  The three chips (``blue'', ``green'', and ``red'')
  are shown with wavelength increasing to the right across the full APOGEE wavelength range of 1.514--1.696~$\mu$m.  
  The gaps between the chips are slightly larger than as displayed in this image.
  Each fiber is
  imaged onto several pixels (vertically).    Note the vertical
  series of points from sky lines in each fiber, and the horizontal
  spectra of faint stars and sky fibers. 
(bottom) Expanded view of the central 18 fibers and central 6~nm of each chip.
}
\label{fig:apogee_2d}
\end{figure*}

\subsection{The APOGEE Instrument and Observations}
\label{sec:apogee_instrument}

The APOGEE spectrograph measures 300 spectra in a single observation:
roughly 230 science targets, 35 on blank areas of sky to measure
sky emission, and 35 hot, blue stars to calibrate atmospheric
absorption.  This multiplexing is accomplished using the same aluminum 
plates and fiber optic technology as have been used for the optical
spectrograph surveys of SDSS.  Each plate corresponds to a specific
patch of sky, and is pre-drilled with holes corresponding to the sky
positions of objects in that area, meaning that each area requires one
or more unique plates. 

The APOGEE spectrograph uses three detectors to cover the $H$-band
range, ``blue'': 1.514--1.581~$\mu$m, ``green'': 1.585--1.644~$\mu$m,
and ``red'': 1.647--1.696~$\mu$m.  There are two gaps, each a few nm
wide, in wavelength in the spectra.  The spectral line spread function
spans 1.6--3.2
pixels per spectral resolution element FWHM, increasing from blue to
red across the detectors. Thus most of the
blue detector is under-sampled.  Figure~\ref{fig:apogee_2d}
shows the results of a typical exposure.  Each observation consists of
at least one ``AB'' pair of exposures for a given pointing on the sky,
with the detector array mechanically offset by 0.5 pixels along the dispersion
direction between the two exposures.  
This well-controlled sub-pixel dithering allows the
derivation of combined spectra with approximately twice the sampling
of the individual
exposures.  
Thus the combined spectra are properly sampled, including
all wavelengths from the blue detector.  
The actual line spread function as a
function of wavelength is provided as a Gauss-Hermite function for
each APOGEE spectrum in DR10.

\begin{figure}
\plotone{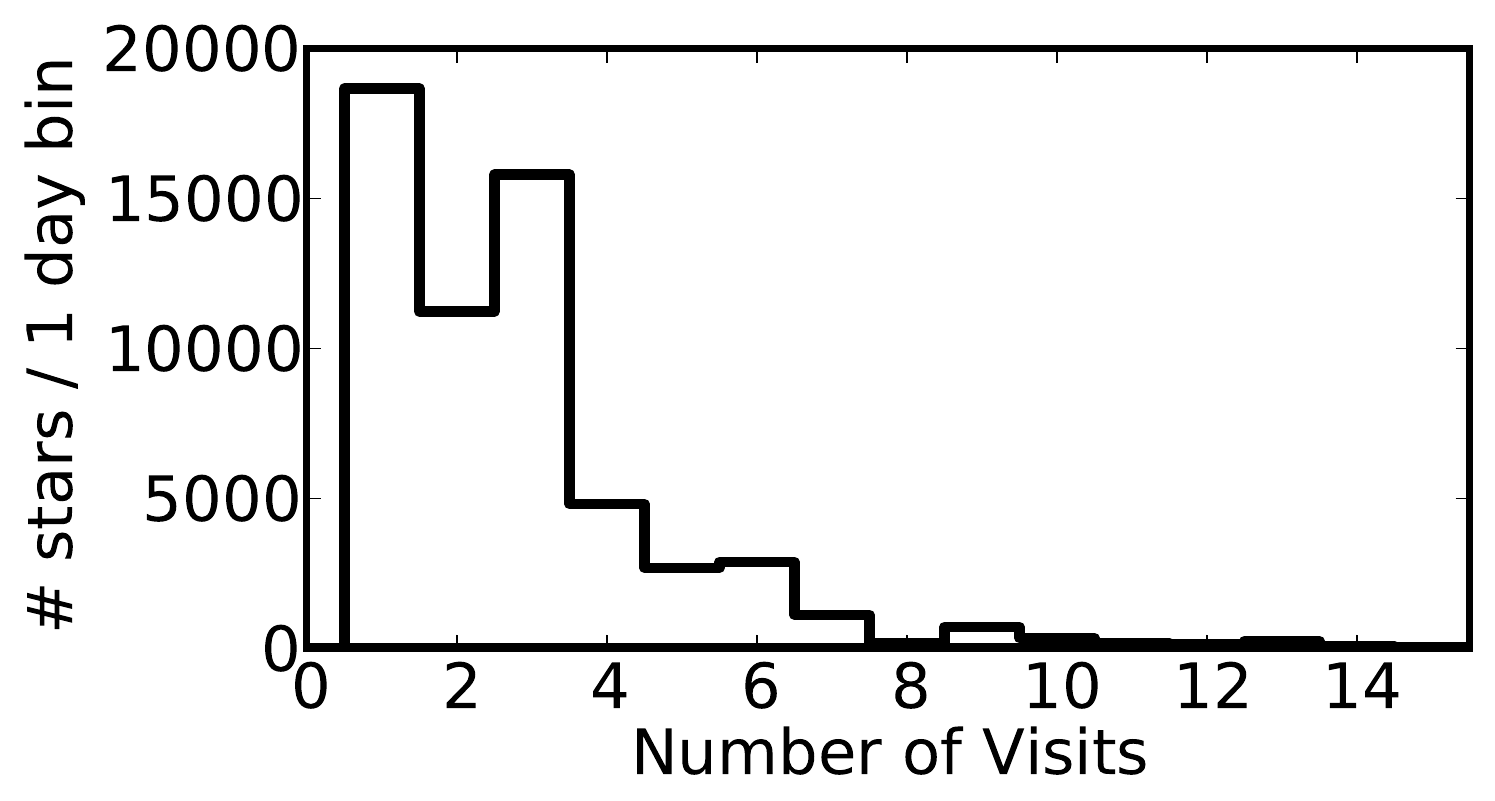}
\caption{The distribution of number of spectroscopic visits for APOGEE stars included in
  DR10.  While the bulk of stars have three or fewer visits, they may
  have reached our spectral S/N requirement if they are bright enough;
  see Figure~\ref{fig:apogee_snr_mag}.
}
\label{fig:apogee_nvisit}
\end{figure}

\begin{figure}
\plotone{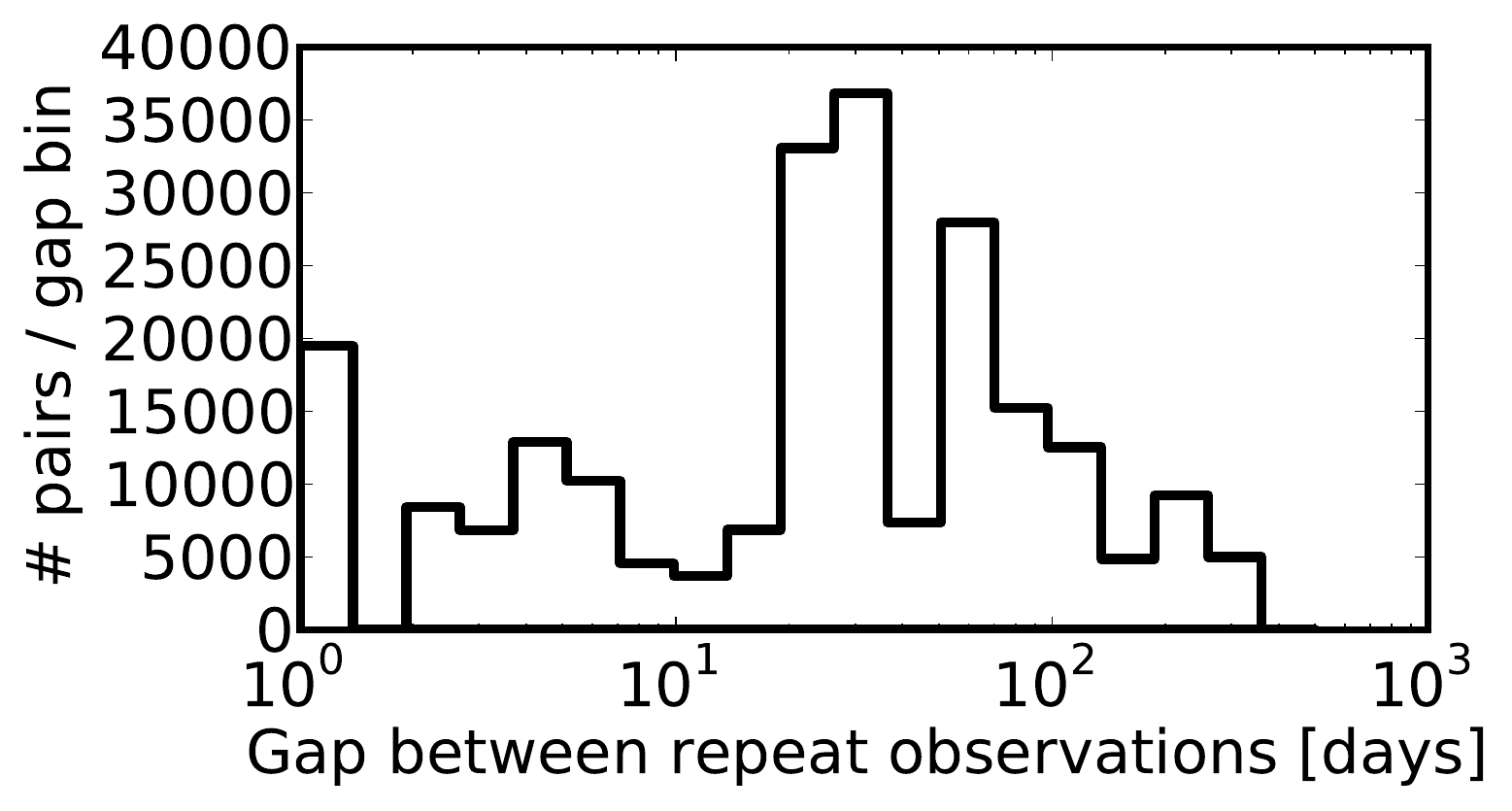}
\caption{The distribution of time between visits for APOGEE stars,
  useful for determining the sensitivity to radial velocity variations
  due to binarity.
This quantity is the absolute value of the time difference for all
unique pairs of visits for each star.  The most prominent peaks
are at one and two months.  
}
\label{fig:apogee_cadence}
\end{figure}

A typical observation strategy is two ``ABBA'' sequences.  Each
sequence consists of four 500-second exposures to reach the target S/N
for a given observation.  The combination of all ``AB'' or ``BA'' pairs for a
given plate during a night is called a ``visit.''  The visit is the
basic product for what are considered individual spectra for APOGEE
(although the spectra from the individual exposures are also made available).
While the total exposure time for a visit is 4,000 seconds ($2\times4\times500$~seconds), due to the
varying lengths of night and other scheduling issues, we often
gathered more or less than the standard two ``ABBA'' sequences on a given
plate in a night.  
APOGEE stars are observed over multiple visits (the goal is at least
three visits) to achieve the planned
S/N. Figure~\ref{fig:apogee_nvisit} shows the distribution of
the number of visits for stars included in DR10; presently, most stars have
three or fewer visits, but this distribution will broaden with
the final data release.  These visits are separated across different
nights and often different seasons, allowing us to look for radial
velocity variability due to binarity on a variety of timescales.  The distribution of
time intervals between visits is shown in
Figure~\ref{fig:apogee_cadence}, with peaks at one and two lunations
(30 and 60 days).

Each visit is uniquely identified by the plate number and MJD of the
observation.  Plates are generally re-plugged between observations, so
while ``plate+MJD+fiber'' remains a unique identifier in APOGEE
spectra as it is in optical SDSS spectra,
``plate+fiber'' does not refer to the same object across all visits.
The spectra from all visits are co-added to produce the aggregate
spectrum of the star.
The final co-added spectra are processed by the stellar
parameters pipeline described in Section~\ref{sec:apogee_aspcap}.

\begin{figure}
\plotone{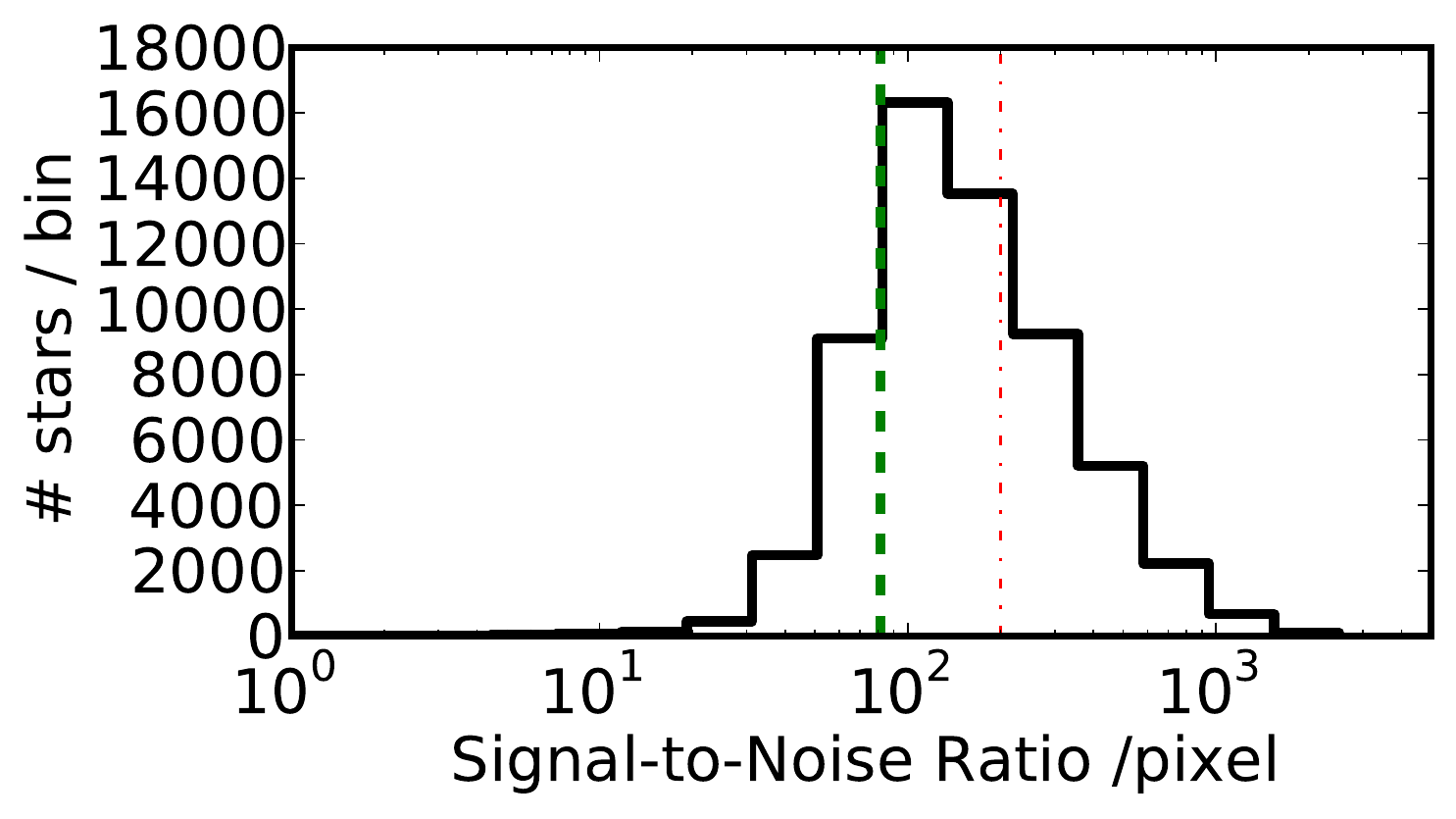}
\caption{Reported S/N per pixel of APOGEE DR10
  co-added stellar spectra. 
Repeated observations imply that there is a 
practical limit of $\rm S/N\sim200$ in the co-added spectra, shown as
the dot-dashed line. 
The dashed line denotes the goal of $\rm S/N \sim 100$ per half-resolution
element, corresponding to $\rm S/N \sim 80$ per pixel in the co-added spectra.
}
\label{fig:apogee_snr}
\end{figure}

\begin{figure}
\plotone{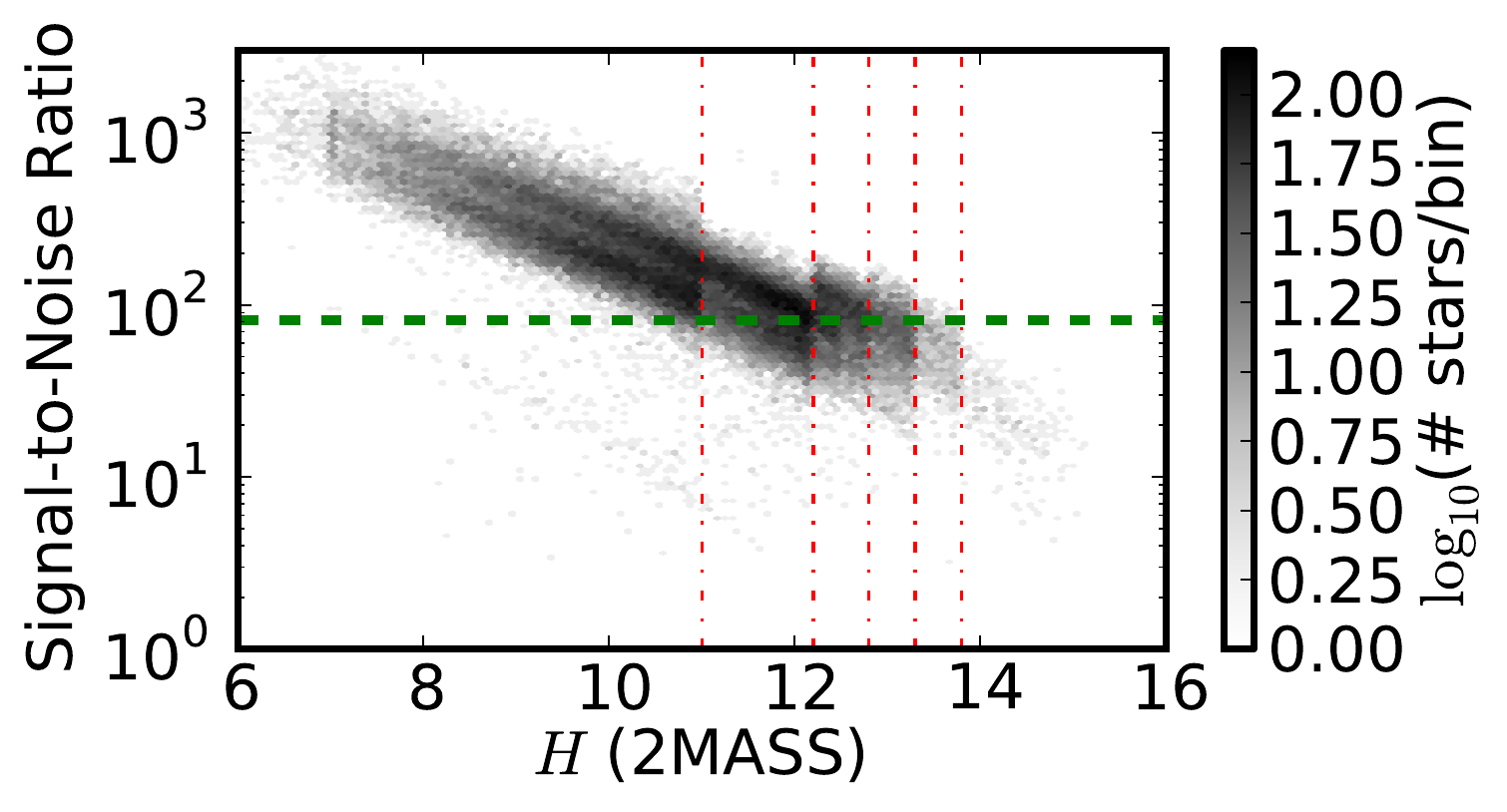}
\caption{S/N per pixel of spectra of stars as a function of their apparent $H$-band magnitude (density is on a log scale).
The vertical dot-dashed lines indicate the magnitude limits for stars
at each value of the final number of visits: 1, 3, 6, 12, 24 visits for $H=11.0$, 12.2, 12.8, 13.3, and 13.8~mag.
The horizontal dashed line denotes the target $\rm S/N \sim 100$ per half-resolution
element, corresponding to $\rm S/N \sim 80$ per pixel in the co-added spectra.
}
\label{fig:apogee_snr_mag}
\end{figure}

  The aim is for a final co-added spectrum of each star with a
  S/N of $>100$ per half-resolution element.\footnote{This is a refinement from the less stringent goal of S/N$>100$ per full-resolution element given in \citet{Eisenstein11}.}  
Figures~\ref{fig:apogee_snr} and
\ref{fig:apogee_snr_mag} show the distribution of S/N; not
surprisingly, S/N is strongly correlated with the brightness of the
star.  The DR10 data include some stars that have yet to receive their full
complement of visits and thus have
significantly lower quality spectra.  Future data releases will
include additional visits for many stars, leading to an increase in
total co-added S/N as well as more refined stellar parameters.

The APOGEE plates are drilled with the same plate-drilling machines
used for BOSS, and the plate numbers are sequential.  This scheme
means that the BOSS and APOGEE plate numbers are interleaved and
that no plate number is assigned to both a BOSS and APOGEE plate.

The quality of the APOGEE commissioning data (that taken prior to 2011
Aug 31) is lower than the survey data, due to optical distortions and
focus issues that were resolved before the official survey was
started.  The biggest difference lies in the ``red'' chip, which has
significantly worse spectral resolution in the commissioning data than
in the survey data.  
Because of this degradation, the data were not under-sampled, and 
spectral dithering was not done during commissioning.  

Many of the targets observed in commissioning were selected in the
same way as those observed during the survey
(Section~\ref{sec:apogee_targeting}), though several 
test plates were designed with different criteria to test
the selection algorithms (e.g., without a color limit or with large numbers of
potential telluric calibration stars). Total exposure times for the
commissioning plates were similar to those of the survey plates. 
Because the spectral resolution of commissioning data is worse, it cannot
be analyzed using ASPCAP with the same spectral libraries with which the
survey data are analyzed. As a result, DR10 does not release any
stellar parameters other than radial velocities for commissioning
data; subsequent releases may 
include stellar parameters for APOGEE commissioning derived using
appropriately matched libraries and/or with only a subset of the spectral range. 

\subsection{APOGEE Main and Ancillary Targets}
\label{sec:apogee_targeting}

APOGEE main targets are selected from 2MASS data \citep{Skrutskie06} 
using apparent magnitude limits to meet the S/N goals and a dereddened
color cut of $(J-\Ks)_0>0.5$~mag to select red giants 
in multiple components of the Galaxy: the disk, bulge, and halo.  
This selection results in a sample of objects that are predominantly
red giant stars with $3500<T_{\rm eff}<5200$~K and $\log{g}<3.5$
(where $g$ is in cm~s$^{-2}$ and the logarithm is base 10).  
Fields receiving three visits have a magnitude limit of $H=12.2$; the
deepest plates with 24 visits go to $H=13.8$.  

APOGEE has also implemented a number of ancillary programs to pursue
specific investigations enabled by its unique instrument.   
The selection of the main target sample and the ancillary programs,
together with the bit flags that can be used to identify why
an object was targeted for spectroscopy, are described in detail in
\citet{Zasowski13}. In DR10, APOGEE stars are named based on a
slightly shortened version of their 2MASS ID (e.g.,
``2M21504373+4215257'' is stored for the formal designation ``2MASS
21504373+4215257''). 
A few objects that don't have 2MASS IDs are designated as ``AP'',
followed by their coordinates.  

APOGEE targets were chosen in a series of fields designed to sample a
wide range of Galactic environments (Figure~\ref{fig:apogee_skydist}):
in the halo predominantly at high latitudes, in the disk, in the
central part of the 
Milky Way (limited in declination), as well as
special targeted fields overlapping the Kepler survey
(Section~\ref{sec:apogee_apokasc}), and a variety of open and globular
clusters with well-characterized metallicity in the literature.  

\begin{figure}
\plotone{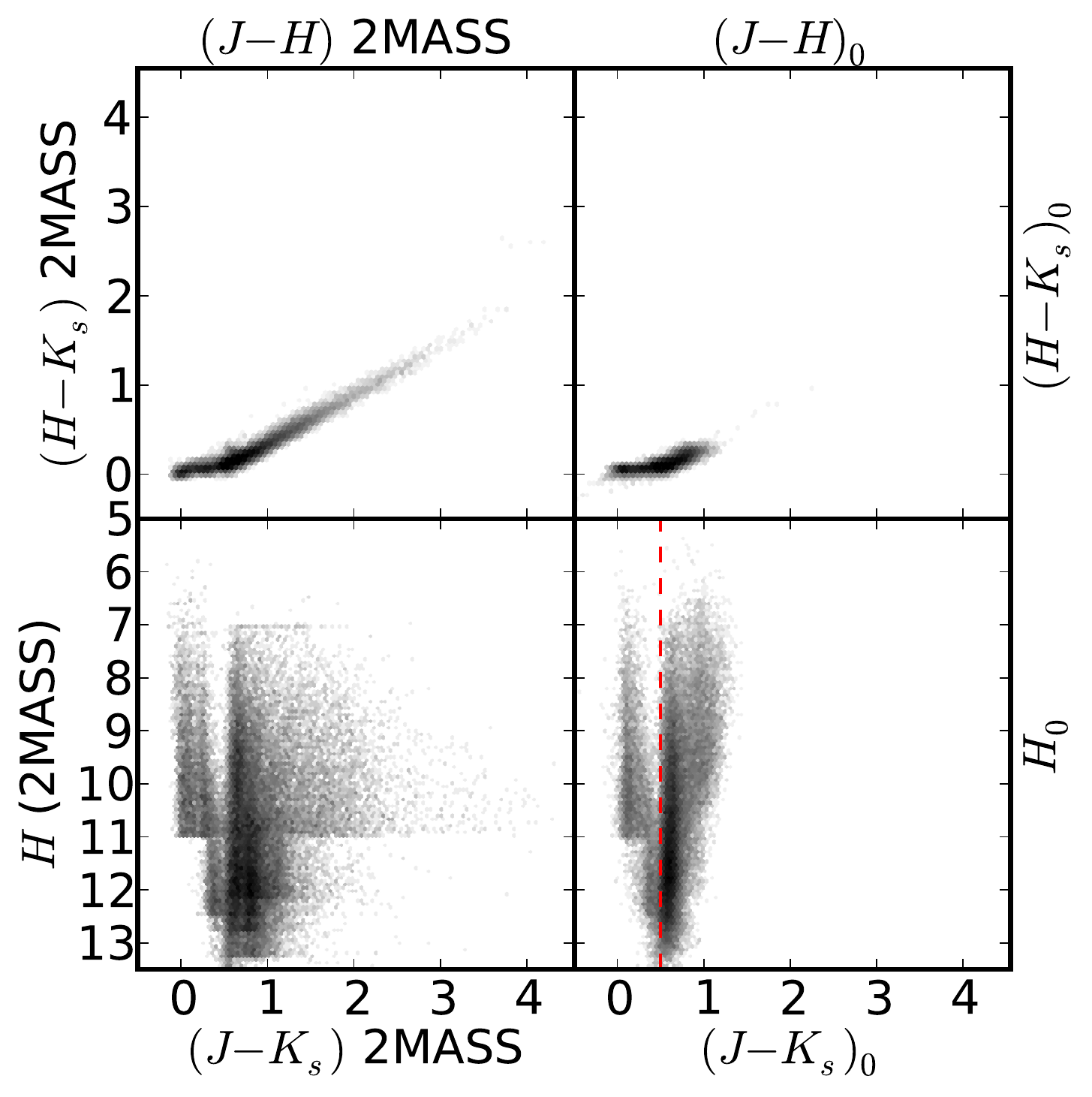}
\caption{Two-dimensional histogram of the APOGEE DR10
  stars in (top) 2MASS $JH\Ks$ color space; and (bottom) 2MASS $H$
  vs. $J-\Ks$.  The left column shows observed magnitudes and colors
  from 2MASS, 
  while the right column has been dereddened based on $H-4.5$~$\mu$m
  color as in \citet{Zasowski13}.  The vertical dashed line at
  $(J-K_s)_0 = 0.5$ shows the selection of the main APOGEE red giant sample;
  bluer objects include telluric calibration stars, data taken during
  commissioning, and ancillary program targets.  The grey scale is
  logarithmic in number of stars.
}
\label{fig:apogee_colors}
\end{figure}

The effects of Galactic extinction on 2MASS photometry can be quite
significant at low Galactic latitude.  We correct for this using 
the {\it Spitzer} IRAC GLIMPSE survey~\citep{Benjamin03,Churchwell09}
and the Wide-field Infrared Survey Explorer (WISE; \citealt{Wright10}) $\lambda=4.5$~$\mu$m data following the
Rayleigh-Jeans Color Excess Method described in \citet{Majewski11} and
\citet{Zasowski13} using the color extinction curve from
\citet{Indebetouw05}.  
Figure~\ref{fig:apogee_colors} shows the measured and
reddening-corrected $JH\Ks$ color-color and magnitude-color diagrams
for the APOGEE stars included in DR10.  

In regions of high interstellar extinction, even intrinsically blue
main sequence stars can be reddened enough to overlap the nominal red
giant locus.  Dereddening these apparent colors allows us
to remove these dwarfs with high efficiency from the final targeted
sample.  However, G and K dwarfs cannot be distinguished
from red giants on the basis of their dereddened broadband colors,
with the
result that a fraction of the APOGEE sample is composed of such
dwarfs. In the disk they are expected to comprise less than 20\% of
the sample, and this appears to be validated by our analysis of the
spectra. Disk dwarfs are expected to be a larger contaminant in halo
fields, so in many of these, target selection was supplemented by
Washington and intermediate-band
DDO51 photometry \citep{Canterna76,Clark79,Majewski00}
using the 1.3-m telescope of the U.S.
Naval Observatory, Flagstaff Station.  Combining this with 2MASS
photometry allows us to distinguish dwarfs
and giants (see \citealt{Zasowski13} for details).

Exceptions to the $(J-\Ks)_0 > 0.5$~mag color limit that appear in
DR10 include the telluric calibration stars, early-type stars targeted
in well-studied open clusters, stars observed on commissioning plates
that did not employ the color limit, and stars in sparsely populated
halo fields where a bluer color limit of $(J-\Ks)_0 > 0.3$~mag was employed to ensure
that all fibers were utilized.  Ancillary program targets
may also have colors and magnitudes beyond the limits of APOGEE's
normal red giant sample.

\subsection{APOKASC}
\label{sec:apogee_apokasc}

Non-radial oscillations are detected in virtually all red giants
targeted by the {\it Kepler} mission \citep{Borucki10,Hekker11}, and
the observed frequencies are sensitive diagnostics of basic stellar
properties such as mass, radius, and age \citep[for a review,
  see][]{Chaplin13}.  Abundances and surface gravities measured from
high-resolution spectroscopy of these same stars are an important
test of stellar evolution models, and allow observational
degeneracies to be broken. 

With this in mind, the ``APOKASC'' collaboration was formed between 
\mbox{SDSS-III} and the {\it Kepler} Asteroseismology Science
Collaboration (KASC) to analyze APOGEE spectra for 
$\sim10,000$ stars in fields observed by the {\it Kepler} telescope 
(see Figure~\ref{fig:apogee_skydist}).  
The joint measurement of masses,
radii, ages, evolutionary states, and chemical abundances for all
these stars will enable significantly enhanced investigations of
Galactic stellar populations and fundamental stellar physics.

DR10 presents 4,204 spectra of 2,308 stars of the anticipated final
APOKASC sample.  Asteroseismic data from the APOKASC collaboration were used to
calibrate the APOGEE spectroscopic surface gravity results for all
APOGEE stars presented in DR10 \citep{Meszaros13}.  A joint
asteroseismic and spectroscopic value-added catalog will be 
released separately (M.~Pinsonneault et al., 2014, in preparation).  

\subsection{APOGEE Data Analysis}
\label{sec:apogee_analysis}

The processing of the two-dimensional spectrograms and extraction of
one-dimensional co-added spectra will be fully 
described in D.~Nidever et al. (2014, in preparation).  We provide here a brief summary to
help the reader understand how individual APOGEE exposures are
processed.  A 500-second APOGEE exposure actually consists of a series
of non-destructive readouts every 10.7 seconds that result in a
three-dimensional data cube.  The first step in processing is to
extract a two-dimensional image from a combination of these
measurements.  After dark current subtraction, the ``up-the-ramp'' values for
each pixel are fit to a line to derive the count rate for that pixel. 
Cosmic rays create characteristic jumps in the ``up-the-ramp'' signal
that are easily recognized, removed, and flagged for future
reference.  The count rate in each pixel is multiplied by the exposure
time to obtain a two-dimensional image.  These two-dimensional images
are then dark-subtracted and flat-fielded.  One-dimensional spectra
are extracted simultaneously for the entire set of 300 fibers based on
wavelength and profile fits from flat-field calibration images.
Both the flat-field response and spectral traces are very 
stable due to the controlled environment of the APOGEE instrument,
which has been under vacuum and at a uniform temperature
continuously since it was commissioned. 
Wavelength calibration is performed using emission lines from thorium-argon and uranium-neon hollow cathode lamps.  
The wavelength solution is then adjusted from the reference lamp calibration on an exposure-to-exposure basis using the location of the night sky lines.

The individual exposure spectra are then corrected for telluric
absorption and sky emission using the sky spectra and telluric calibration star
spectra, and combined accounting for the dither offset between each
``A'' and ``B'' exposure. This combined visit spectrum
is flux-calibrated based on a model of the APOGEE instrument's
response from observations of a blackbody source.  The spectrum is
then scaled to match the 2MASS measured apparent $H$-band magnitude. 
A preliminary radial velocity is measured after matching the visit spectrum
to one from a pre-computed grid of synthetic stellar spectra, and is stored
with the individual visit spectrum.   

\begin{figure*}
\plottwo{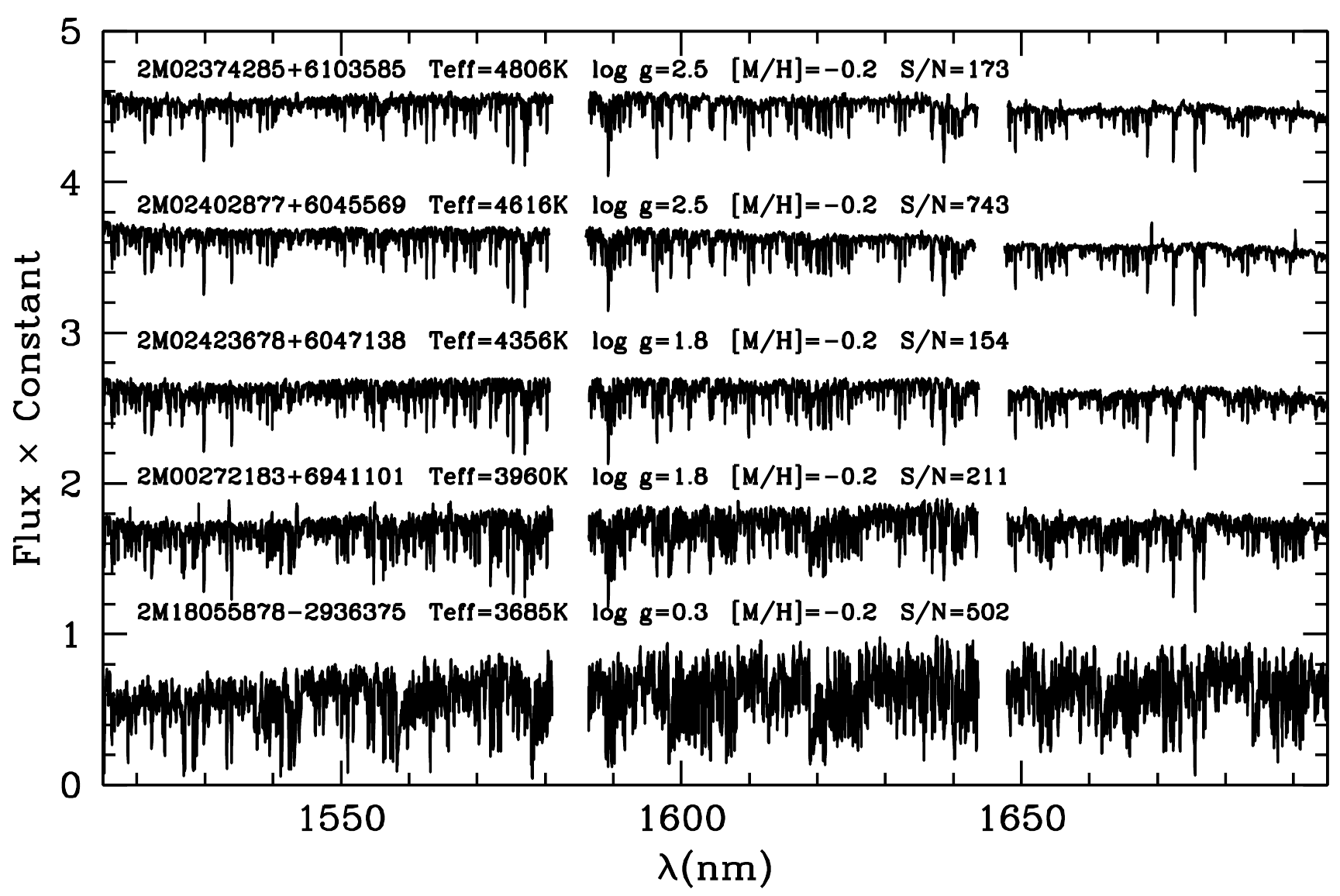}{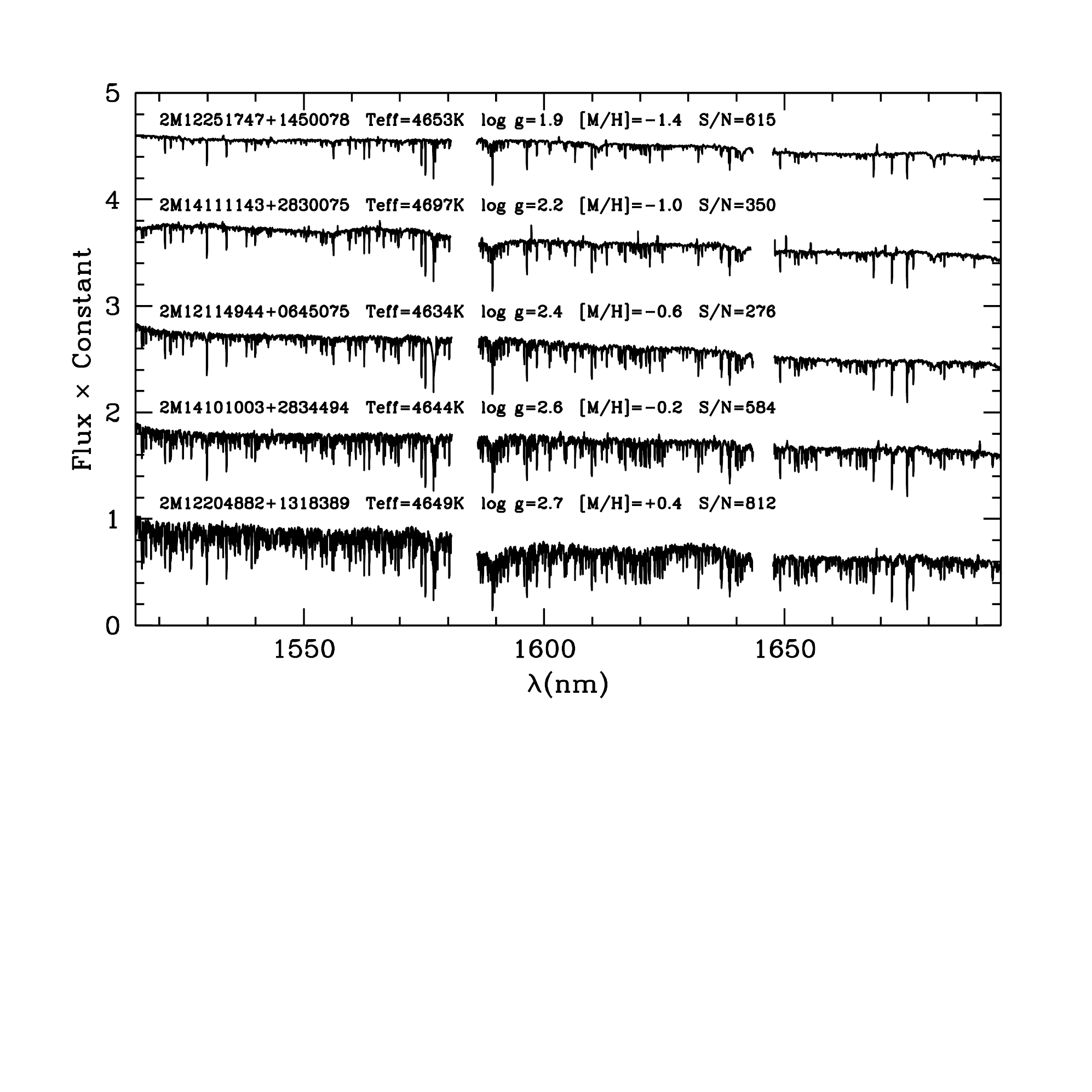}
\caption{Typical APOGEE spectra at high S/N. (left) Spectra of stars with 
  $5000$~K$>T_{\rm eff}>3750$~K at constant [M/H]$=-0.2$ (a
  characteristic [M/H] for the sample).  The trend in line intensity from top
  to bottom is driven by decreasing $T_{\rm eff}$ (which is strongly
  correlated 
  with $\log{g}$ -- see Figure~\ref{fig:apogee_starparams}).
(right) Spectra of stars with $\rm -1.4<$[M/H]$<+0.4$ at constant
  $T_{\rm eff}\sim 4650$~K (a characteristic $T_{\rm eff}$ for the
  sample).   The trend of increasing absorption lines in the spectra
  from top to bottom is driven by the increasing [M/H]. 
  All of
  these spectra have a reported S/N of at least 200 per co-added re-sampled
  pixel: each of the observed absorption lines in the spectra are real features
  of the observed stars.  The apparent emission lines are actually residuals
  from the incomplete subtraction of airglow lines. 
}
\label{fig:apogee_spectra}
\end{figure*}

In addition to the individual visit spectra, the APOGEE software
pipeline coadds the spectra from different visits to the same field,
yielding a higher S/N spectrum of each object.
Figure~\ref{fig:apogee_spectra} shows examples of high S/N co-added
flux-calibrated 
spectra from APOGEE for stars with a range of $T_{\rm eff}$ and with a
range of [M/H].  A final and precise determination of the relative radial
velocities on each visit is determined from cross-correlation of each
visit spectrum with the combined spectrum; the velocities are put on an
absolute scale by cross-correlating the combined spectrum with the
best-matching spectrum in a pre-computed synthetic grid.
The combined spectra are output on a rest-wavelength scale with
logarithmically spaced pixels with approximately three pixels per spectral resolution element.

\subsection{Issues with APOGEE Spectra}
\label{sec:apogee_caveats}

Users should be aware of several
features and potential issues with the APOGEE data. This is the first data
release for APOGEE; the handling of some of these issues by the
pipelines may be improved in subsequent data releases.

Many of these issues are documented in the data by the use
of bitmasks that flag various conditions. For the APOGEE spectral
data, there are two bitmasks that accompany the main data products
Each one-dimensional extracted spectrum includes a signal,
uncertainty, and mask arrays.  The mask array is a bitmask,
\code{APOGEE\_PIXMASK}\footnote{\url{http://www.sdss3.org/dr10/algorithms/bitmask\_apogee\_pixmask.php}},
that flags data-quality conditions that affect a given pixel.  A
non-zero \code{APOGEE\_PIXMASK} value for a pixel indicates a potential
data-quality concern that affects that pixel.  Each stellar-parameters
analysis of each star is accompanied by a single bitmask,
\code{APOGEE\_STARFLAG}\footnote{\url{http://www.sdss3.org/dr10/algorithms/bitmask\_apogee\_starflag.php}},
that flags conditions at the full spectrum level.

The most important data-quality features to be aware of include:

{\bf Gaps in the spectra: } There are gaps in the spectra
corresponding to the regions that fall between the three
detectors. There are additional gaps due to 
bad or hot pixels on the arrays.  As multiple dithered exposures are combined to make
a visit spectrum, values from missing regions cannot be used to calculate
the dither-combined signal in nearby pixels; as a result, these nearby
pixels are set to zero and the \code{BADPIX} bit is set for these pixels in
\code{APOGEE\_PIXMASK}.  
Generally, the bad pixels affect neighboring
pixels only at a very low level, and the data in the latter may be usable; in subsequent
data releases, we will preserve more of the data, while continuing to
identify potential bad pixels in the
pixel mask.

{\bf Imperfect night-sky-line subtraction}: The Earth's atmosphere has
strong and variable emission in OH lines in the APOGEE bandpass. At
the location of 
these lines, the sky flux is many times brighter than the stellar flux
for all except the brightest stars. Even if the sky subtraction
algorithm were perfect, the photon noise at the positions of these sky
lines would dominate the signal, so there is little useful information
at the corresponding wavelengths.  The spectra
in these regions can show significant sky line residuals.  These
regions are masked for the stellar parameter analysis so that they do
not impact the results. 
The affected pixels have the \code{SIG\_SKYLINE} bit set in \code{APOGEE\_PIXMASK}.

{\bf Error arrays do not track correlated errors}: APOGEE spectra from
an individual visit are made by combining multiple individual
exposures taken at different dither positions. Because the dithers
are not spaced by {\em exactly} 0.5 pixels, there is some correlation
between pixels that is introduced when combined spectra are produced.
The error arrays for the visit spectra do not include information
about these correlations. In the visit spectra, these correlations are generally small because
the dither mechanism is generally quite accurate. However, when multiple visit
spectra are combined to make the final combined spectra, they must
be re-sampled onto a common wavelength grid, taking into account the
different observer-frame velocities of each individual visit. This re-sampling
introduces significant additional correlated errors between adjacent pixels that are
also not tracked in the error arrays.

{\bf Error arrays do not include systematic error floors}:  
The errors that are reported for each spectrum are derived based
on propagation of Poisson and readout noise.  However, based on
observations of bright hot stars, we believe
that other, possibly systematic, uncertainties currently limit APOGEE observations
to a maximum S/N per half resolution element of $\sim 200$.  The error arrays published in DR10 currently report
the estimated errors without any contribution from a systematic
component. However, for the ASPCAP analysis, we impose an error
floor corresponding to 0.5\%  of the continuum level.

{\bf Fiber crosstalk}: While an effort is made not to put faint stars
adjacent to bright ones on the detector to avoid excessive spillage of
light from one to the other, this occasionally occurs.  We
flag objects (in \code{APOGEE\_STARFLAG}) with a
\code{BRIGHT\_NEIGHBOR} flag if an adjacent star is $>10$ times
brighter than the object, and with a \code{VERY\_BRIGHT\_NEIGHBOR}
flag if an adjacent star is $>100$ times brighter; in the latter case,
the individual spectra 
are marked as bad and are not used in combined spectra.

{\bf Persistence in the ``blue'' chip}:  There is a known ``superpersistence'' in
1/3 of the region of the ``blue'' APOGEE data array, and to a lesser extent in some regions of 
the ``green" chip, whereby some of the charge from previous exposures
persists in subsequent exposures.  Thus the values read out in these
locations depend on the previous exposure history for that chip.
The effect of superpersistence can vary significantly, but residual
signal can amount to as much as 10--20\% of the signal from previous
exposures. 
The current pipeline does not
attempt to correct for this effect; any such correction is likely to be rather
complex. For the current release, pixels known to be affected by
persistence are flagged in \code{APOGEE\_PIXMASK} at three different levels
(\code{PERSIST\_LOW, PERSIST\_MEDIUM, PERSIST\_HIGH}). Spectra that have significant
numbers of pixels ($>20$\% of total pixels) that fall in the persistence
region have comparable bits set in the \code{APOGEE\_STARFLAG} bitmask to warn that
the spectra for these objects may be contaminated. In a few cases, the
effect of persistence is seen dramatically as an elevated number of counts
in the blue chip relative to the other arrays; these are flagged as \code{PERSIST\_JUMP\_POS}
in \code{APOGEE\_STARFLAG}.
We are still actively investigating the effect of persistence on
APOGEE spectra and derived stellar parameters, and are working on
corrections that we intend to implement for future data releases.

\subsection{APOGEE Stellar Parameter and Chemical Abundances Pipeline (ASPCAP)}
\label{sec:apogee_aspcap}

The ultimate goal of APOGEE is to determine the effective temperature,
surface gravity, overall metallicity, and detailed chemical abundances for
a large sample of stars in the Milky Way.  
Stellar parameters and chemical abundances are extracted from
the continuum-normalized co-added APOGEE
spectra by comparing with synthetic spectra calculated using
state-of-the-art model photospheres \citep{Meszaros12}  
and atomic and molecular line opacities (Shetrone et al., in
preparation). 

\begin{figure}
\plotone{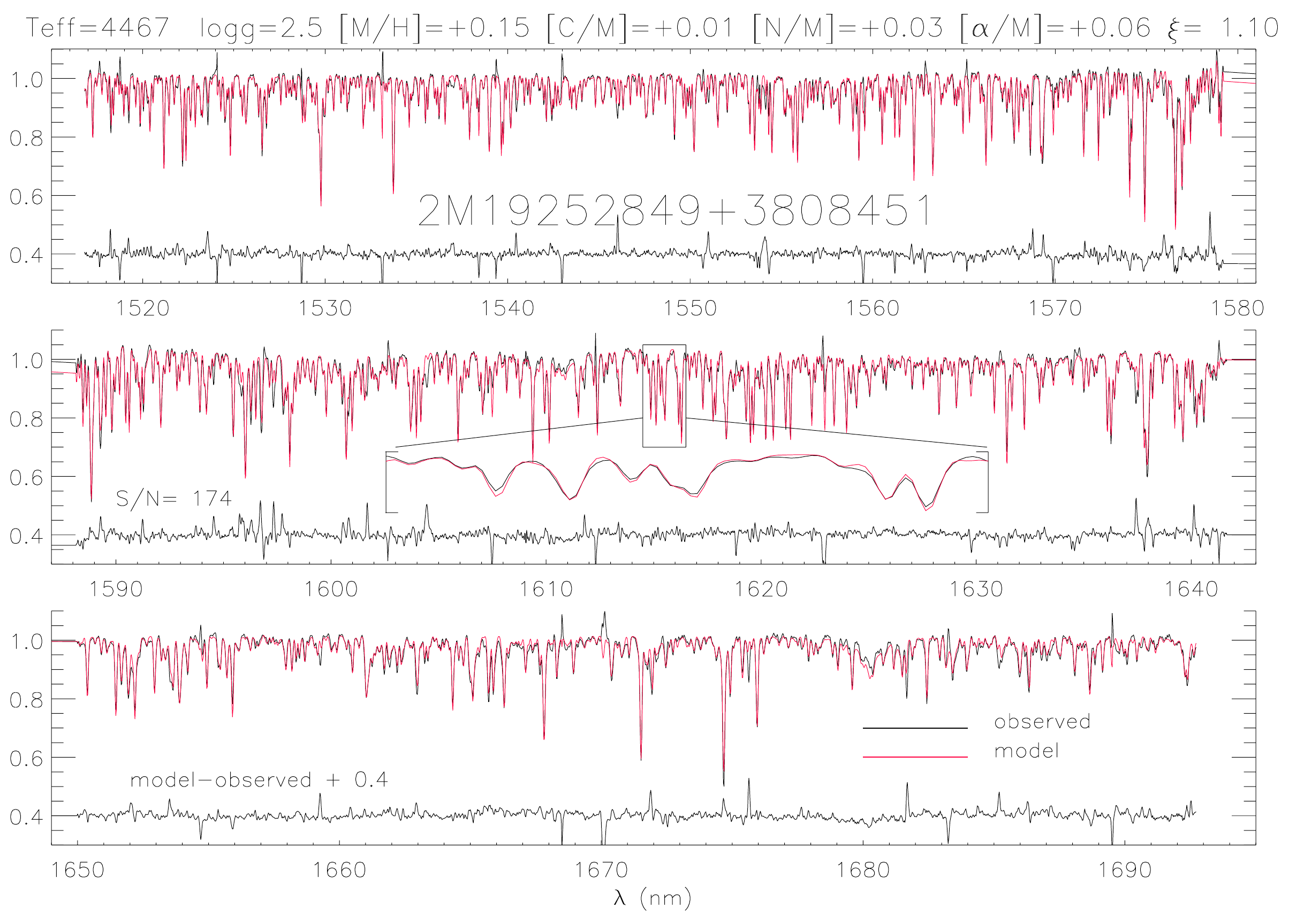}
\caption{(upper lines) An example ASPCAP fit (red) to a typical APOGEE co-added
  stellar spectrum (black).  
(lower lines) Residual of the ASPCAP model fit compared to the data (offset from zero by +0.4 units for clarity of presentation).
(inset) Zoom on a region showing the high resolution of the actual data.
The H-band spectrum contains a wealth of information about the
elemental abundances and stellar parameters of the star.  The high
resolution and high S/N of APOGEE spectra allow these atmospheric
properties to be measured for the entire APOGEE
sample.  
}
\label{fig:apogee_aspcap}
\end{figure}

Analysis of high-resolution spectra is traditionally done by hand.
However, given the sheer size of APOGEE's spectral database, automatic
analysis methods must be implemented.  For that purpose, ASPCAP
searches for the best fitting spectrum through $\chi^2$ minimization
within a pre-computed multi-dimensional grid of synthetic spectra, 
allowing
for interpolation within the grid.
The output parameters of the analysis are effective temperature
($T_{\rm eff}$), surface gravity ($\log g$), metallicity ([M/H]), and
the relative abundances of $\alpha$ elements ([$\alpha$/M])\footnote{The relative $\alpha$-element abundance is labeled \code{ALPHAFE} in the DR10 tables and files, but it is more accurately the ratio of the $\alpha$ elements to the overall metallicity, [$\alpha$/M].}, 
carbon
([C/M]), and nitrogen ([N/M]).  
The micro-turbulence quoted in the DR10 results is not an independent quantity, but is instead calculated directly from the value of $\log g$.
Figure~\ref{fig:apogee_aspcap} shows an example
ASPCAP fit to an APOGEE spectrum of a typical star.
ASPCAP will be fully described in an
upcoming paper (A.~Garc\'\i{}a P\'erez et al., 2014, in preparation). 

Chemical composition parameters are defined as follows.  The abundance
of a given element $\rm X$ is defined relative to solar values in the standard way:
\begin{equation}
[{\rm X/H}] = \log_{10}(n_{\rm X}/n_{\rm H})_{\rm star} -
\log_{10}(n_{\rm X}/n_{\rm H})_\Sun\, ,
\end{equation}
where $n_{\rm X}$ and $n_{\rm H}$ are respectively the numbers of atoms of element
$\rm X$ and hydrogen, per unit volume, in the stellar photosphere.  The
parameter [M/H] is defined as an overall metallicity scaling, assuming
the solar abundance pattern.  The deviation of the abundance of
element $\rm X$ from that pattern is given by
\begin{equation}
\rm [X/M] = [X/H] - [M/H] \, .
\end{equation}
The $\alpha$ elements considered in the APOGEE spectral libraries are
O, Ne, Mg, Si, S, Ca, and Ti, and [$\alpha$/H] is defined as an
overall scaling of the abundances of those elements, where they are
assumed to vary together while keeping their relative abundances fixed
at solar ratios.  For DR10, we allow four chemical composition parameters to
vary: the overall metallicity, and the abundances of $\alpha$
elements, carbon, and nitrogen.  
Carbon, nitrogen, and
oxygen contribute significantly to the opacity in APOGEE spectra of cool giants,
particularly in the form of molecular lines due to OH, CO, and CN.

\subsubsection{Parameter Accuracies}

\citet{Meszaros13} have compared the outputs of ASPCAP to
stellar parameters in the literature for
stars targeted by APOGEE in open and globular clusters spanning a wide
range in metallicity.   
These comparisons uncovered small systematic differences between
ASPCAP and literature results, which are mostly based on
high-resolution optical spectroscopy.  These differences are not
entirely understood yet, and we hope they will be corrected in future
data releases.  In the meantime, calibrations have been derived to
bring APOGEE and literature values into agreement.
With these offsets in place, the APOGEE metallicities are
accurate to within 0.1~dex for stars of $\rm S/N > 100$ per half-resolution
element that lie within a strict range of
$T_{\rm eff}$, $\log g$, and [M/H].
Based on observed scatter in the ASPCAP calibration clusters, we estimate that the internal precision of the APOGEE measurements is 0.2~dex for $\log g$, 150~K for $T_{\rm eff}$, and 0.1~dex for [$\alpha$/M]
 \citep[see][for details]{Meszaros13}.  

Because most of the observed cluster stars are giants, the applied
calibration offsets only apply to giants. The parameters
of dwarfs are generally accurate enough to determine that they
are indeed higher surface gravity stars, but otherwise their
parameters are likely to be more uncertain: one reason for this
is that rotation is likely to be important for a larger fraction
of these stars, and the effects of rotation are not currently included
in our model spectral libraries.

APOGEE mean values 
per cluster of [$\alpha$/M] are in good agreement with those in the literature.
However, there are systematic correlations
between [$\alpha$/M] and both [M/H] and $T_{\rm eff}$ for stars
outside the range $\rm -0.5 \leq$[M/H]$\leq 0.1$.  Moreover, important
systematic effects may be present in [$\alpha$/M] for stars cooler
than $T_{\rm eff} \sim 4200$~K.
We therefore discourage use of [$\alpha$/M] for stars with $T_{\rm eff}<4200$~K or with [M/H]$<-0.5$ or [M/H]$>+0.1$.

Figure~15 in \citet{Meszaros13} shows the root-mean square scatter in
[$\alpha$/M] for red giants in open and globular clusters, as a
measure of the uncertainty in this parameter.  However, given the
trends in [$\alpha$/M] with other stellar parameters, care should be
taken when estimating the accuracy of [$\alpha$/M].

Comparison with literature values for carbon
and nitrogen abundances shows large scatter and significant systematic
differences.  In view of the relative paucity and uncertainty of
literature data for these elements, more work is needed to understand
these systematic and random differences before APOGEE abundances for
carbon and nitrogen can be confidently adopted in science applications.   

\subsubsection{ASPCAP Outputs}

In DR10, we provide calibrated values of effective temperature, surface
gravity, overall metallicity, and [$\alpha$/M] for giants. In addition,
we provide the raw ASPCAP results (uncalibrated, and thus, to some extent,
unvalidated) for all six parameters for all stars with survey-quality data.
Since commissioning data have lower resolution, different spectral libraries
are needed to derive stellar parameters from them, and therefore ASPCAP results
are not provided for these spectra at this time. For all stars with
ASPCAP results, we also provide information about the quality of the
fit ($\chi^2$) and several bitmasks (\code{APOGEE\_ASPCAPFLAG} and \code{APOGEE\_PARAMFLAG})
that flag several conditions that may cause the results to be less
reliable. Among these conditions are abnormally high $\chi^2$ in the
fit, best-fit parameters near the edges of the searched range,
evidence in the spectrum of significant stellar rotation, and so
on. Users should check the values of these bitmasks before using the ASPCAP parameters.

\begin{figure*}
\plotone{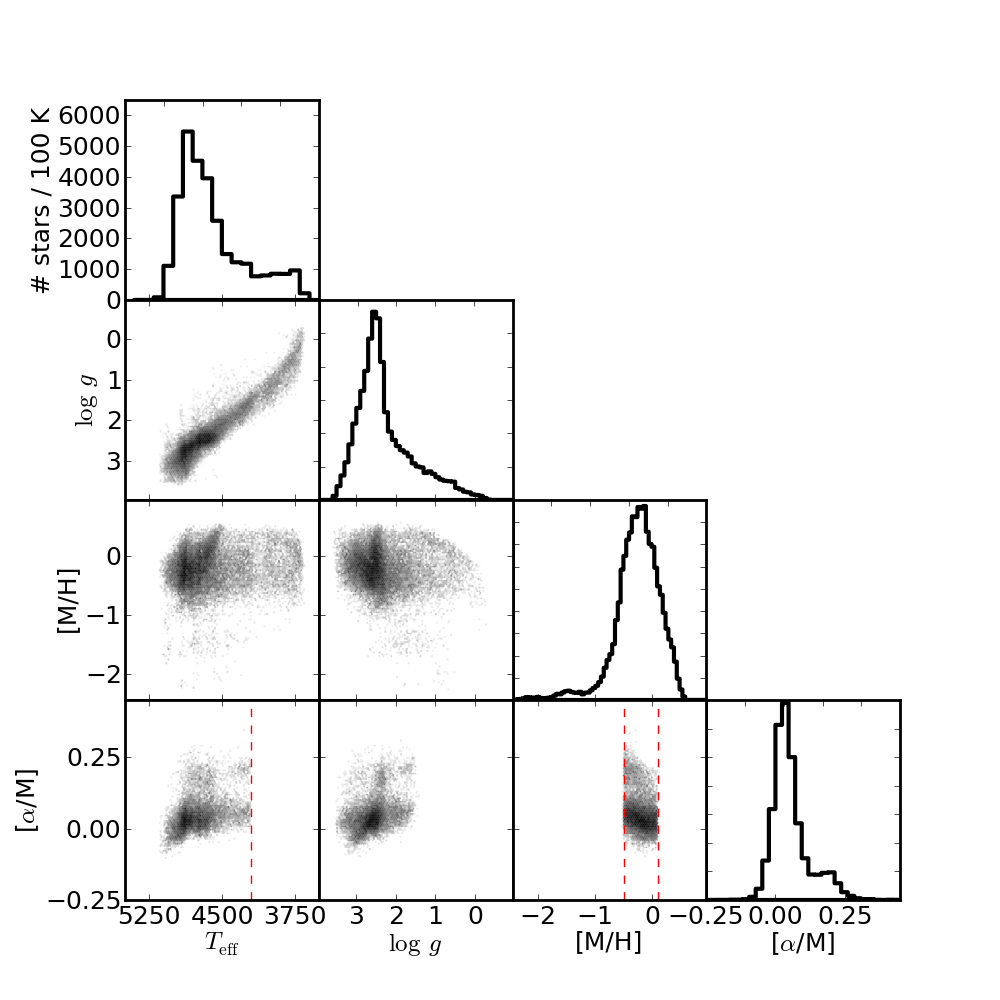}
\caption{The one-dimensional and two-dimensional distributions of APOGEE stellar parameters ---
  temperatures, surface gravities, metallity,
  and [$\alpha$/M] --- 
for all 29,438 APOGEE stars in DR10 which have reliable ASPCAP fits.
The [$\alpha$/M] values are only shown for the 16,066 star subset with
$T_{\rm eff}>4200$~K and $-0.5<$[M/H]$<+0.1$, which is the range for which [$\alpha$/M] values are reliable (limits are indicated by red dashed lines; see Section~\ref{sec:apogee_aspcap} for details).
These distributions show what APOGEE has observed and ASPCAP has analyzed.  They do not represent a fair sample of the underlying Galactic populations.
}
\label{fig:apogee_starparams}
\end{figure*}

Figure~\ref{fig:apogee_starparams} shows the distribution of stellar properties derived by ASPCAP for
stars included in DR10.  
The ASPCAP spectral libraries are currently only calibrated in the
range $3610<T_{\rm eff}<5255$~K.
Thus the reliable ASPCAP $T_{\rm eff}$ reported values lie only in this range,
with a peak at about 4800~K. 
The surface gravity
distribution peaks at $\log g \sim 2.5$, corresponding to red
clump stars, 
and
is strongly correlated with surface temperature. 
The ASPCAP models are calibrated in the range $-0.5<\log{g}<3.6$, which is reflected in the range shown.
Because of the strong concentration of targeted fields to the Galactic
plane (Figure~\ref{fig:apogee_skydist}), the metallicity distribution
peaks just below solar 
levels, 
with a tail extending from ${\rm [M/H]} \sim -0.5$ to below $-2.3$.
The [$\alpha$/M] abundance distribution has both $\alpha$-rich and $\alpha$-poor stars, which reflects the variety of populations explored by APOGEE.

\begin{figure*}
\plotone{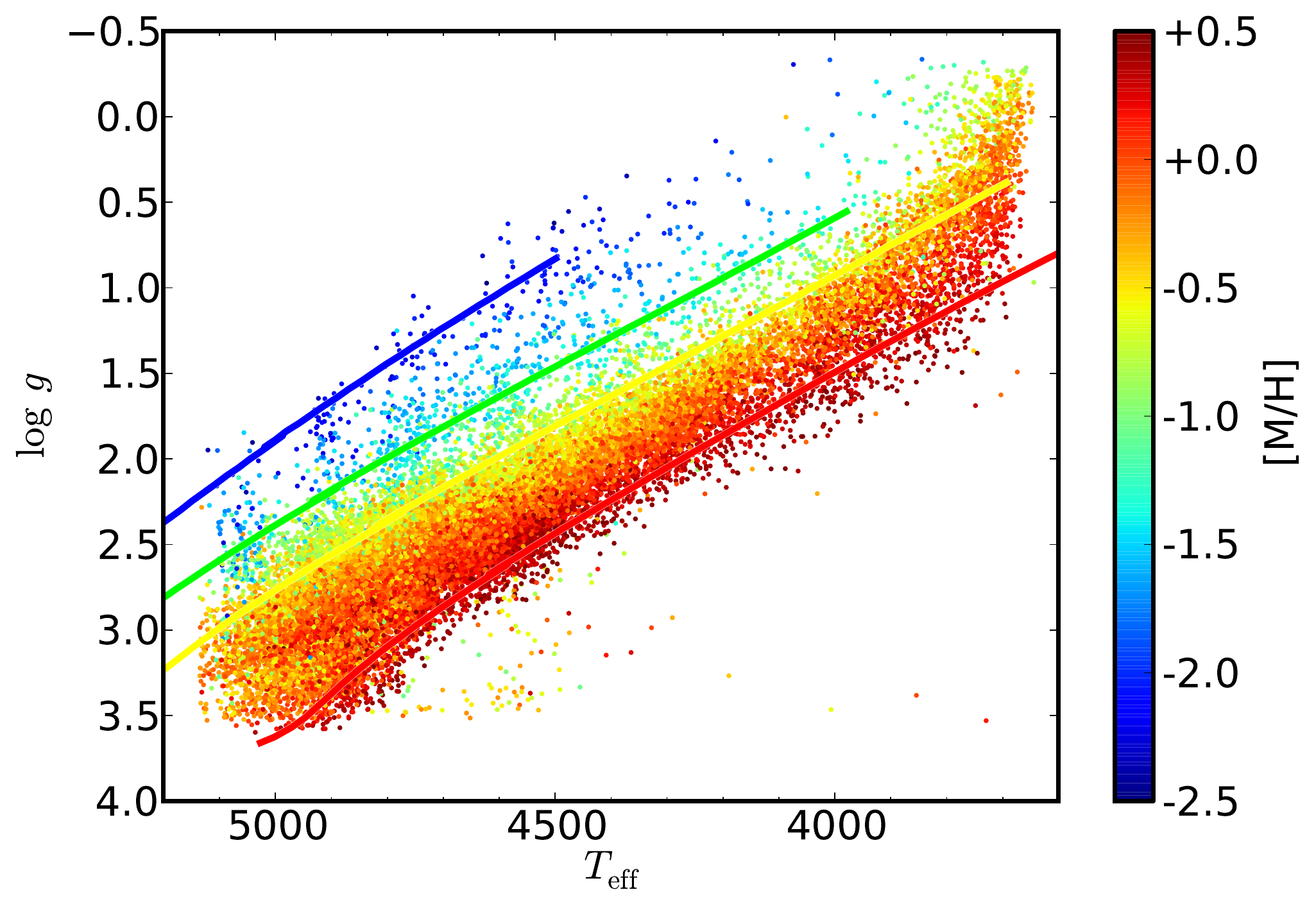}
\caption{ASPCAP $\log{\,g}$ vs. $T_{\rm eff}$ with the points
  color-coded by [M/H].  Overplotted are isochrones for a 4 Gyr
  population of RGB stars with [$\alpha$/Fe]$=0$ from
  \citet{Bressan12} on the same color-coded metallicity scale.
  The isochrones are for [M/H] = $-1.9, -1.0, -0.58$, and
  +0.14 from left to right.} 
\label{fig:apogee_logg_teff_iso}
\end{figure*}

Figure~\ref{fig:apogee_logg_teff_iso} shows the excellent agreement of the ASPCAP $\log{\,g}$, $T_{\rm eff}$, and [M/H] values with the isochrone models of \citet{Bressan12}.

\subsection{APOGEE Data Products}
\label{sec:apogee_data}

The APOGEE data as presented in DR10 are available as the individual
500-second spectra taken on a per-exposure basis (organized both by
object and by plate+MJD+fiber), as combined co-added spectra on a
per-object basis, and as continuum-normalized spectra 
used by the APOGEE pipeline (ASPCAP) when it computes stellar properties (Section~\ref{sec:apogee_aspcap}). 
The individual raw exposure files, processed spectra, and combined
summary files of stellar parameters are provided as
FITS\footnote{\url{http://fits.gsfc.nasa.gov/}} files~\citep{FITS} 
through the DR10 Science Archive Server (SAS).
The DR10 Catalog Archive Server (CAS) provides the basic stellar parameters
(including the radial velocity) 
from the APOGEE spectra on a per-visit (SQL table \code{apogeeVisit}) and a co-added star basis (SQL table \code{apogeeStar}). 
The ASPCAP results are provided
in the SQL table \code{aspcapStar}; the covariances between these
parameters are given in a separate table, \code{aspcapStarCovar}.

To allow one to recreate the sample selection, all of the parameters
used in selecting APOGEE targets are provided in DR10 in the SQL table
\code{apogeeObject}. 

Example queries for APOGEE data using the CAS are provided as part of the DR10 web documentation\footnote{\url{http://www.sdss3.org/dr10/irspec/catalogs.php\#examples}}.

\section{The Baryon Oscillation Spectroscopic Survey (BOSS)}
\label{sec:boss}

An overview of the BOSS survey is presented in detail in
\citet{Dawson13}, and the 
instrument is described in \citet{Smee13}.  BOSS is obtaining spectra
of 1.5 million galaxies \citep{DR9}, and 150,000 quasars with
redshifts between 2.15 and 3.5 \citep{Ross12}, selected from 10,000
deg$^2$ of SDSS imaging data.  The large-scale distribution of
galaxies and the structure in the quasar Lyman $\alpha$ forest, allow
measurements of the baryon oscillation signature as a function of
redshift \citep{Anderson12, Anderson13, Busca13}.
In addition, about 5\% of the
fibers are devoted to a series of ancillary programs with a broad
range of science goals (see the Appendix of \citealt{Dawson13}).  

DR9 included about 830,000 BOSS spectra over 3275~deg$^2$ from 1.5 years of observation; DR10 adds
an additional 679,000 spectroscopic observations over 3100~deg$^2$ from an additional year of observation that featured unusually good weather at APO.
The quality of the data is essentially unchanged from DR9.  The
spectra cover the wavelength range 3650--10,400\AA, with a resolution
of roughly $R\sim1800$.  The S/N is of course a strong
function of magnitude, but at a model magnitude of $i=19.9$, the
magnitude limit of the CMASS galaxy sample \citep[see][]{Dawson13,DR9}, the typical 
median S/N per pixel across the spectra is about 2.  The 
majority of these spectra are of adequate quality for classification
and measurement of a redshift; 6\% of the galaxy target
spectra and 12\% of the quasar target spectra are flagged by
the spectroscopic pipeline \citep{Bolton12} as having uncertain
classification.  These numbers are significantly higher than they were
for SDSS-I/II, as the targets are quite a bit fainter, but they remain
small enough for quantitative analysis of the samples (especially with
visual inspections of the quasar targets; see \citealt{Paris12}).

\begin{figure}
\plotone{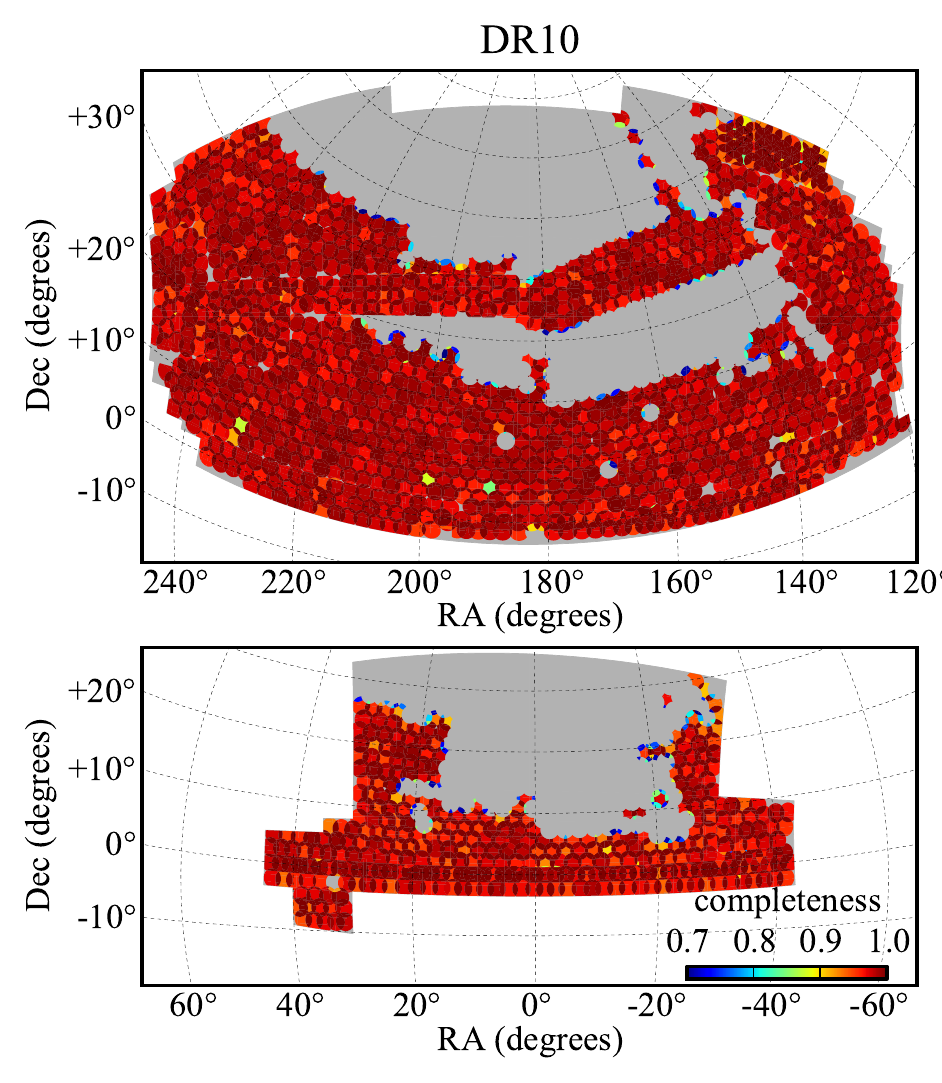}
\caption{BOSS DR10 spectroscopic sky coverage in the Northern Galactic
Cap (top) and Southern Galactic Cap (bottom).  The grey region is
the coverage goal for the final survey, totaling 10,000~deg$^2$.
The color coding indicates the fraction of CMASS galaxy targets
that receive a fiber; the fact that no two fibers can be placed
closer than $62$\arcsec\ on a given plate reduces the average completeness
to 94\%.  Note the higher completeness on the Equator in the
Southern Galactic Cap (Stripe~82) where the plates are tiled with more overlapping
area to recover collided galaxies.
\label{fig:boss_coverage}
}
\end{figure}

  Figure~\ref{fig:boss_coverage} shows the sky coverage of the BOSS
  spectroscopic survey in more detail than in
  Figure~\ref{fig:skydist_boss}. The tiling of the individual circular
  plates is visible in this completeness map of the CMASS galaxy
  sample.  Because of the finite extent of the cladding around fibers,
  no two fibers can be placed closer than $62$\arcsec, meaning that
  spectroscopy will be only about 94\% complete in regions covered by
  only a single plate.

\begin{figure}
\plotone{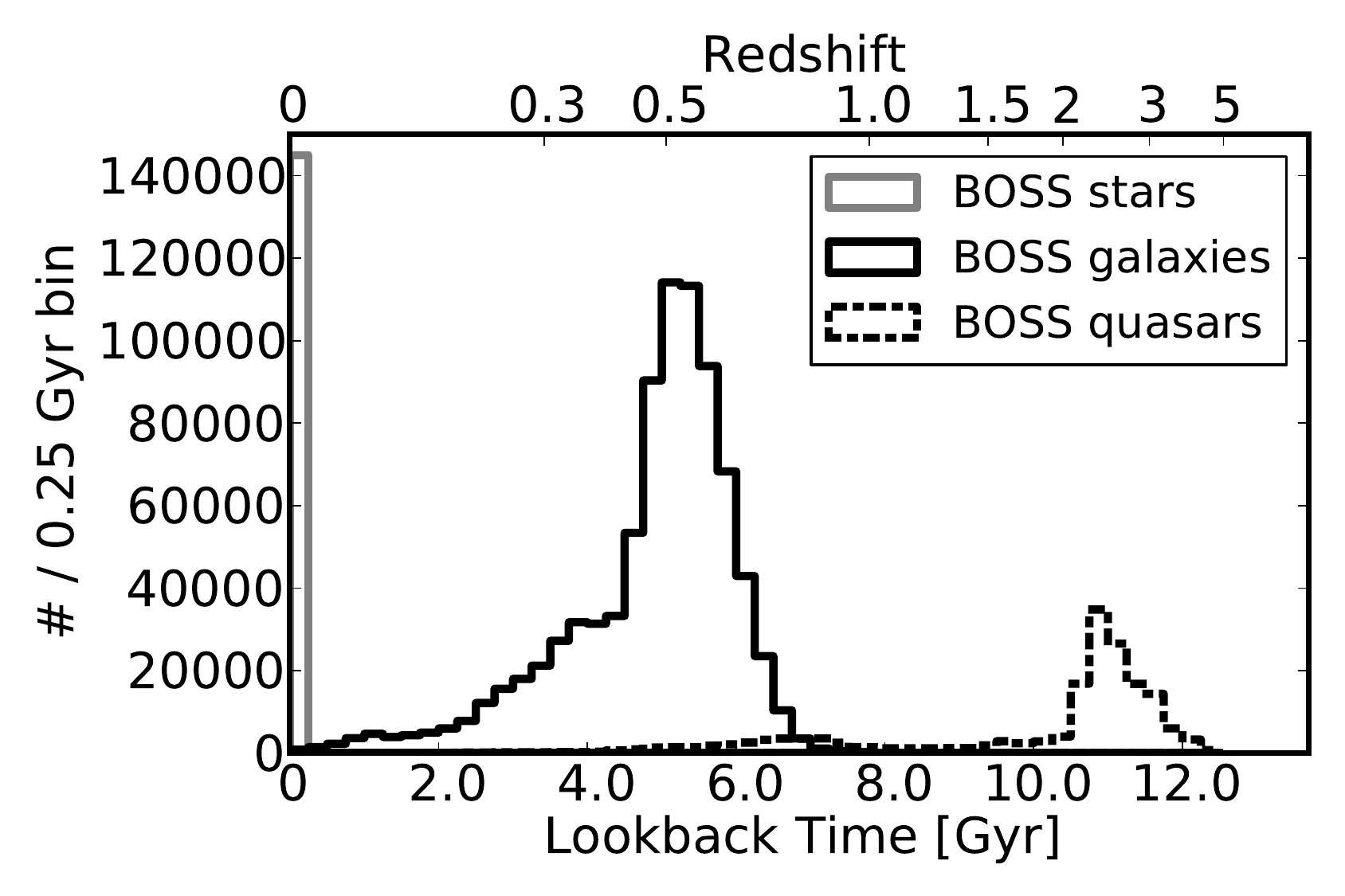}
\caption{
The distribution of BOSS DR10 spectroscopic objects versus lookback time for the
144,968 unique stars; 859,322 unique galaxies; and
166,300 unique quasars.
Lookback time is based on the observed redshift under the assumption of
a $\Lambda$CDM cosmology \citep{Komatsu11}. 
This figure is nearly identical to 
the equivalent for DR9 (Figure~3 of \citet{DR9}),
scaled by a factor of 1.8.
}
\label{fig:boss_star_galaxy_qso}
\end{figure}

 Figure~\ref{fig:boss_star_galaxy_qso} shows the distribution of DR10 BOSS
 spectroscopy as a function of lookback time, or equivalently
 redshift.  The galaxy distribution peaks at a redshift of 0.5 (about
 5.5 Gyr ago), with
 very few galaxies above redshift 0.7.  By design, the majority of quasars lie
 between redshifts 2.15 and 3.5, as this is the range in which
 the Lyman $\alpha$ forest enters the BOSS spectral coverage.  

\begin{figure}
\plotone{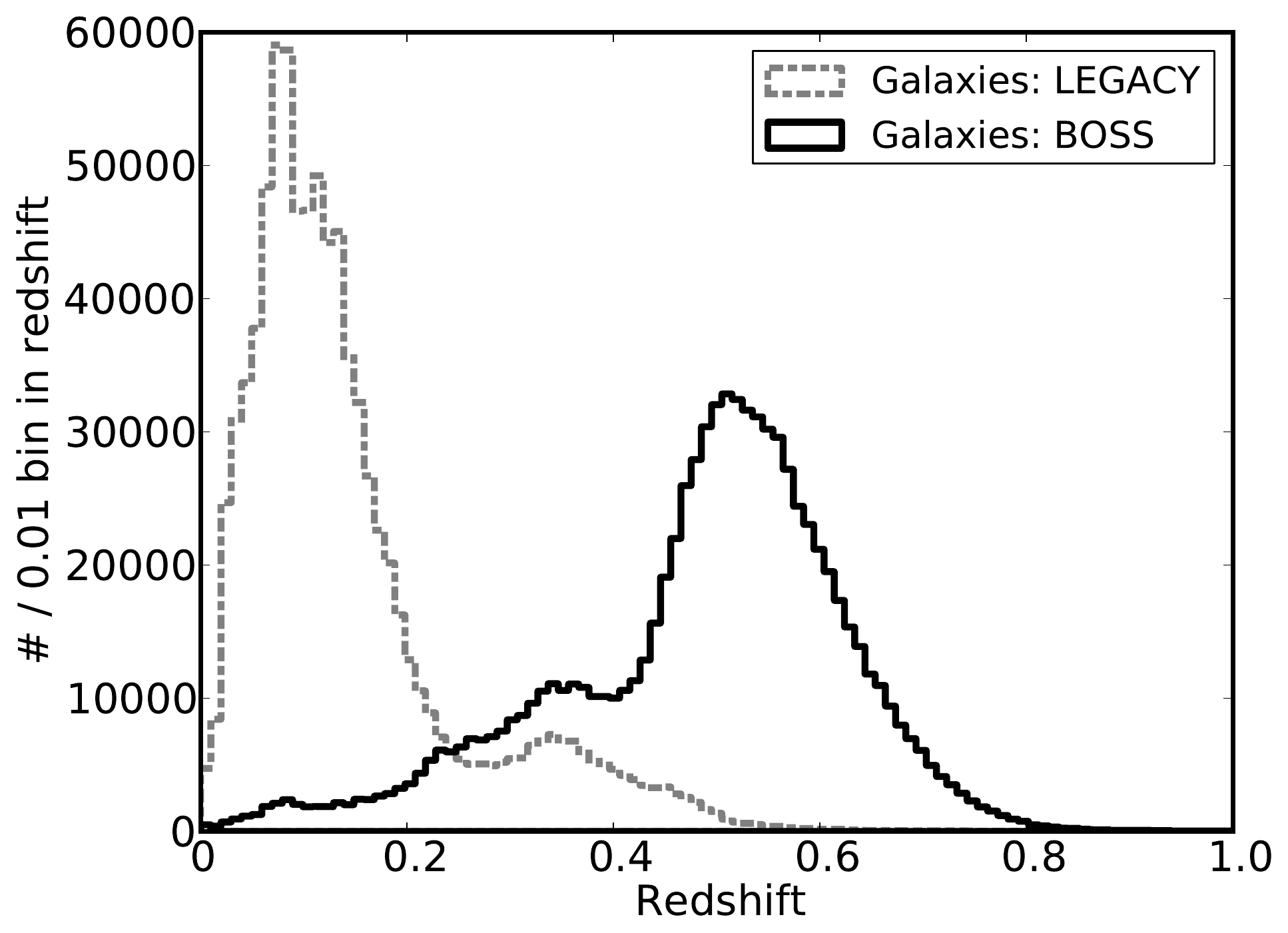}
\plotone{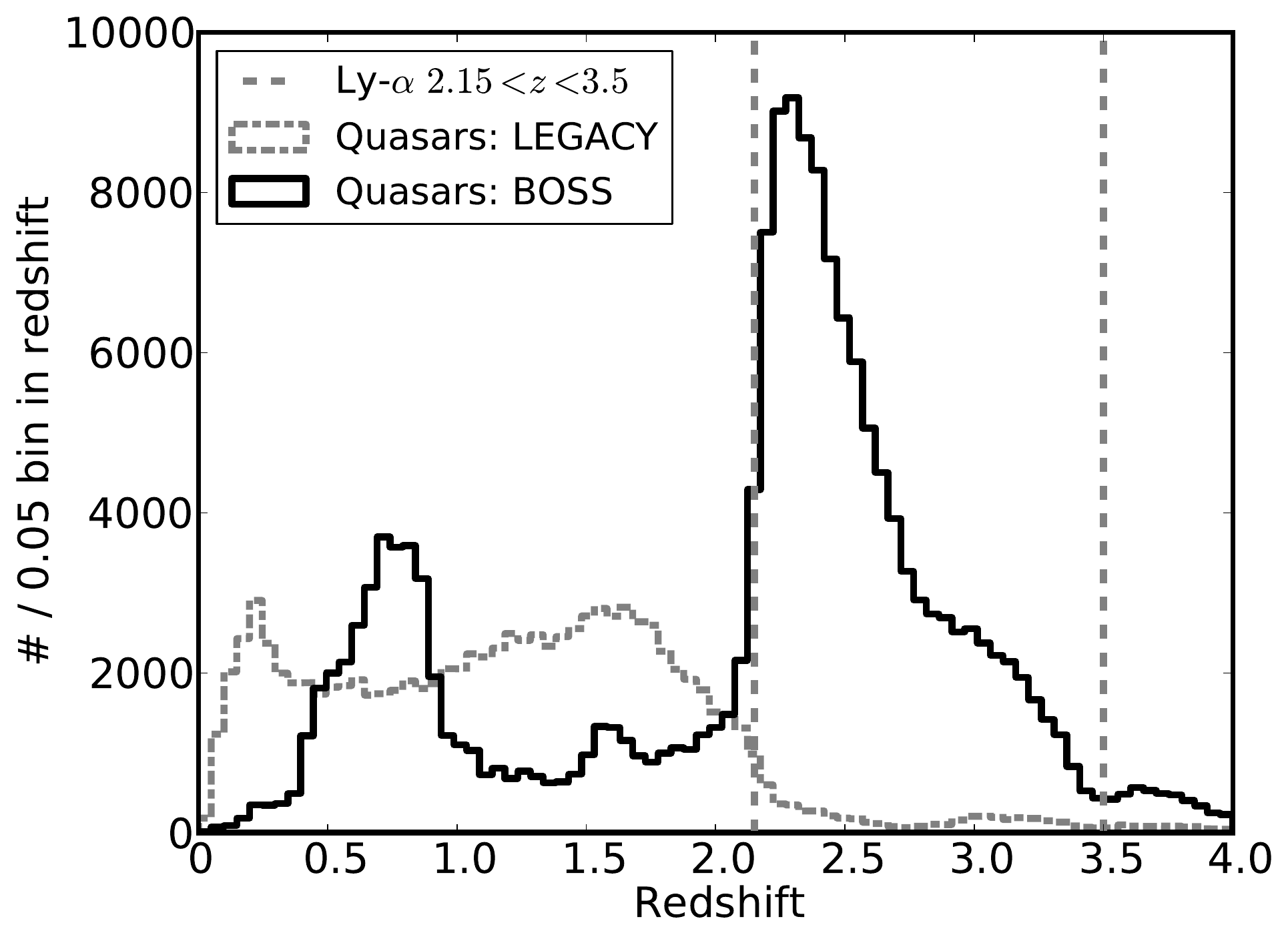}
\caption{$N(z)$ of \mbox{SDSS-III} BOSS spectra in DR10 compared to that of the
  SDSS-I/II Legacy spectra for galaxies (top) and quasars (bottom).
}
\label{fig:boss_galaxy_qso_vs_sdss}
\end{figure}

These distributions are shown in more detail in
Figure~\ref{fig:boss_galaxy_qso_vs_sdss}, which compares the redshift
distributions of galaxies and quasars to those from the SDSS-I/II
Legacy survey.  The SDSS-I/II galaxy survey includes a magnitude-limited
sample with median redshift $z \approx 0.10$ \citep{Strauss02} and a
magnitude- and color-selected sample of luminous red galaxies
extending to beyond $z = 0.4$ \citep{Eisenstein01}.  The SDSS-I/II
quasar survey \citep{Richards02,DR7Q} selects quasars at all redshifts and is
flux-limited at magnitudes significantly brighter than BOSS; the bulk
of the resulting quasar sample lies below $z = 2$.  The BOSS DR10 galaxy
sample is roughly the same size as the full DR7 Legacy galaxy sample
(at almost five times the median redshift) and the BOSS DR10 quasar
sample is significantly larger than its Legacy counterpart.  DR10
includes about 60\% of the full BOSS footprint, so DR12, the final
SDSS-III data release, will be roughly 50\% larger. 

In what follows, Section~\ref{sec:boss_wiseboss} describes a new
quasar
target class for quasars selected using WISE data,
Section~\ref{sec:boss_pipeline} describes minor updates to the 
BOSS spectroscopic pipeline in DR10, and Section~\ref{sec:boss_galparams}
discusses additions to measurements of parameters from galaxy
spectra.

\subsection{A New Quasar Target Class in DR10}
\label{sec:boss_wiseboss}

\citet{Ross12} describe the quasar target selection used in BOSS.  
DR10 includes one new quasar 
target class, \code{BOSS\_WISE\_SUPP}, which uses photometry from SDSS and
WISE to
select $z>2$ quasars that the standard BOSS quasar target selection
may have missed, and to explore the properties of quasars selected in the
infrared.

These objects were required to have detections in the 3.6~$\mu$m,
4.5~$\mu$m, and 12~$\mu$m bands, and to be point sources in SDSS
imaging.  They were selected with the following color cuts: 
\begin{equation} 
(u-g)>0.4 \ {\rm and}\ (g-r)<1.3 .
\end{equation}
The requirement of a 12~$\mu$m detection removes essentially all 
stellar contamination, without any WISE color cuts.  

There are 5,007 spectra from this sample in DR10, 
with a density of $\sim1.5$/deg$^2$ over the $\sim 3,100$~deg$^2$ of
new area added by BOSS in DR10.  Almost 3000 of these objects are
spectroscopically confirmed to be quasars, with redshifts up to
$z=3.8$.  Nine-hundred ninety-nine of these objects have $z > 2.15$.  

Given the use of WISE photometry in target selection, we have imported
the WISE All-Sky Release catalog \citep{Cutri12} into the SDSS Catalog
Archive Server (CAS), and performed an astrometric cross-match with
$4$\arcsec\ matching radius with the SDSS catalog objects.  
We find no systematic shift between the WISE and SDSS
astrometric systems; $4$\arcsec\ extends well into the tail of the match
distance distribution.  The results of this matching are also 
available as individual files in the Science Archive
Server (SAS).

\subsection{Updates to BOSS Data Processing}
\label{sec:boss_pipeline}

We have become aware of transient hot columns on the spectrograph
CCDs.  Because fiber traces lie approximately along columns, a bad
column can adversely affect a large swath of a given spectrum.  With
this in mind, unusual-looking spectra associated with fibers 40, 556,
and 834 and fibers immediately adjacent should be treated with
suspicion; these objects are often erroneously classified as $z>5$
quasars.  We will improve the masking of these bad columns in future
data releases.

We have identified 2748 objects with spectra whose astrometry is unreliable in the
SDSS imaging due to tracking or focus problems of the SDSS
telescope while scanning.  As a consequence, the fibers may be
somewhat offset from the true position of the object, often missing it
entirely (and thus having a spectrum with no signal).  The redshift
determination of each object 
is accompanied by a warning flag, \code{ZWARNING}, which indicates that
the results are not reliable (Table 2 of \citealt{Dawson13}).  Objects
with bad astrometry are assigned bit 8, \code{BAD\_TARGET} in \code{ZWARNING}.

\subsection{Updates to BOSS Galaxy Stellar Population Parameters}
\label{sec:boss_galparams}

Estimating stellar population properties for galaxies from SDSS
spectra continues to be an active field with different valid
approaches. DR9 included various estimates of stellar
population parameters, including: 
\begin{itemize} 
\item ``Portsmouth'' stellar masses derived from spectroscopic
  redshifts plus the SDSS imaging $ugriz$ \citep{Maraston13}; 
\item ``Portsmouth'' measurements of stellar kinematics and emission-line
  fluxes combined with model spectral fits to the full spectra
  \citep{Thomas13}, and 
\item ``Wisconsin'' principal component analysis 
(PCA) of the stellar populations using fits to the wavelength range
$\lambda=3700$--$5500$~\AA\ \citep{Chen12}.
\end{itemize}
The latter two spectral fits include estimates of stellar velocity dispersions.
These measurements agree with each other and the pipeline estimates of
\citet{Bolton12} within their measurement errors, but slight
systematic offsets remain. For a detailed comparison we refer the
reader to \citet{Thomas13}.

All stellar population calculations use the WMAP7 $\Lambda$CDM
cosmology with $H_0=70$~km/s/Mpc, $\Omega_M=0.274$, and
$\Omega_\Lambda=0.726$\ \citep{Komatsu11}.

In DR9, these models were calculated just for BOSS spectra; in DR10
they are extended to the $\sim 930,000$ galaxy spectra from SDSS-I/II.
The Portsmouth code results in DR10 now also include the full stellar mass
probability distribution function for each spectrum.
The Wisconsin PCA code in DR9 used the stellar population model of
\citet{BruzualCharlot03}. In DR10, we have added
the stellar population synthesis model of
\citet{MarastonStroembaeck11}. 
In addition, the covariance matrix in the flux density in neighboring
pixels due to errors in spectrophotometry has been updated by
using all of the repeat galaxy observations in DR10, rather than the
5,000 randomly selected repeat galaxy observations used in DR9.
This covariance is important in fitting stellar
population models to the spectra.

In DR9 we also provided measurements of emission-line fluxes and
equivalent widths as well as gas kinematics \citep{Thomas13}. However,
the continuum fluxes as listed in the Portsmouth DR9 catalog needed
to be corrected to rest-frame by multiplication by
$1+z$. Consequently, the equivalent widths needed to be divided by
the same factor $1+z$ to be translated into the rest frame.  In
DR10, the continuum fluxes and equivalent widths have these correction
factors applied, and are presented in the rest-frame.

In DR10, we also include results from the Granada Stellar Mass code
(A.~Montero-Dorta et al., 2014, in preparation) based on the publicly available
``Flexible Stellar Population Synthesis'' code of \citet{Conroy09}.
The Granada FSPS product follows a similar spectrophotometric SED
fitting approach as that of the Portsmouth galaxy product, but using
different stellar population synthesis models, with varying star
formation history (based on simple $\tau$-models), 
metallicity and dust attenuation.  The Granada FSPS galaxy product
provides spectrophotometric stellar masses, ages, specific star
formation rates, and other stellar population properties, along with
corresponding errors, for eight different models, which are generated
by applying simple, physically motivated priors to the parent grid.
These eight models are based on three binary choices: (1) including or not
including dust; (2) using the \citet{Kroupa01} vs.\ the \citet{Salpeter55} stellar
initial mass function; and (3) two different configurations for the
galaxy formation time: either the galaxy formed within the first 2 Gyr
following the Big Bang ($z\sim3.25$), or the galaxy formed between the
time of the Big Bang and two Gyr before the observed redshift of the
galaxy.

\section{Data Distribution}
\label{sec:distribution}

All Data Release 10 data are available through data access tools linked
from the DR10 web site.\footnote{\url{http://www.sdss3.org/dr10/}}
The data are stored both as individual files in the Science Archive Server (SAS)
and as a searchable database in the Catalog Archive Server (CAS).
Both of these data servers have front-end web interfaces, called the
``SAS Webapp''\footnote{\url{http://data.sdss3.org/}} and
``SkyServer''\footnote{\url{http://skyserver.sdss3.org/dr10/}}, respectively.
A number of different interfaces are available,
each designed to accomplish a specific task.
\begin{itemize}
\item Color images of regions of the sky in JPEG format (based on the
$g$, $r$ and $i$ images; see \citealt{Lupton04}) can be viewed in a web
browser with the SkyServer Navigate tool.  These are presented at
higher resolution, and with greater fidelity, than in previous
releases.  With DR10 we also include JPEG images of the 2MASS data to
complement the APOGEE spectra. 
\item FITS images can be searched for, viewed, and downloaded through the SAS Webapp.
\item Complete catalog information (astrometry, photometry, etc.) of any imaging
object can be viewed through the SkyServer Explore tool.
\item Individual spectra, both optical and infrared, can be searched for,
viewed, and downloaded through the SAS Webapp.
\item Catalog search tools are available through the SkyServer interface to the CAS, each
of which returns catalog data for objects that match supplied
criteria.  For more advanced queries, a powerful and flexible catalog
search website called ``CasJobs'' allows users to create their own
personalized data sets and
then to modify or graph their data.
\end{itemize}
Links to all of these methods are provided at
\url{http://www.sdss3.org/dr10/data\_access/}.

The DR10 web site also features data access tutorials, a glossary of
SDSS terms, and detailed documentation about algorithms used to
process the imaging and spectroscopic data and select
spectroscopic targets.

Imaging and spectroscopic data from all prior data releases are
also available through
DR10 data access tools, with the sole caveat that the 303 imaging runs
covering the equatorial stripe in the Fall sky (``Stripe 82'') are
only fully provided in
DR7\footnote{\url{http://skyserver.sdss.org/dr7}} -- only the good
quality images are included from Stripe 82 in DR8 and subsequent
releases.

\section{Future}
\label{sec:future}

The \mbox{SDSS-III} project will present two more public data
releases: DR11 and DR12, both to be released in December 2014.
DR11 will include data taken through the summer of 2013.
DR12 will be the final \mbox{SDSS-III} data release and will include
the final data through Summer 2014 from all observations with APOGEE, BOSS, MARVELS, and SEGUE-2.

In 2014 July, operation of the 2.5-m Sloan Foundation Telescope will be taken over
by the next generation of SDSS, currently known as SDSS-IV, which plans to operate
for six years.  
SDSS-IV consists of three surveys mapping the Milky Way Galaxy, the
nearby galaxy population, and the distant universe.   
APOGEE-2 will continue the current APOGEE program of targeting Milky Way stars to study Galactic archaeology and stellar astrophysics.  
It will include a southern component, observing from the 2.5-m du Pont
Telescope at Las Campanas Observatory, Chile, allowing a full-sky view
of the structure of the Milky Way. 
Mapping Nearby Galaxies at APO (MaNGA) will use the BOSS spectrograph in a new mode, bundling fibers into integral field units to observe 10,000 nearby galaxies with spatially resolved spectroscopy. 
MaNGA has already observed a small number of targets using BOSS time
to test its planned hardware
configuration. 
Finally, the Extended Baryon Oscillation Spectroscopic Survey (eBOSS)
will create the largest volume three-dimensional map of the universe
to date, to measure baryon acoustic oscillations and
constrain cosmological parameters in the critical and largely
unexplored redshift range $0.6<z<2.1$.  eBOSS will also obtain spectra
of X-ray sources detected by the eROSITA satellite \citep{eROSITA}, as
well as of variable stars and quasars to understand their physical
nature.  
The SDSS-IV collaboration will continue the production and distribution of cutting-edge and diverse data sets through the end of the decade.

\acknowledgments
\mbox{SDSS-III} Data Release 10 makes use of data products from the Two Micron All Sky Survey, which is a joint project of the University of Massachusetts and the Infrared Processing and Analysis Center/California Institute of Technology, funded by the National Aeronautics and Space Administration and the National Science Foundation.

\mbox{SDSS-III} Data Release 10 makes use of data products from the Wide-field Infrared Survey Explorer, which is a joint project of the University of California, Los Angeles, and the Jet Propulsion Laboratory/California Institute of Technology, funded by the National Aeronautics and Space Administration.

Funding for \mbox{SDSS-III} has been provided by the Alfred P. Sloan Foundation, the Participating Institutions, the National Science Foundation, and the U.S. Department of Energy Office of Science. The \mbox{SDSS-III} web site is http://www.sdss3.org/.

\mbox{SDSS-III} is managed by the Astrophysical Research Consortium for the Participating Institutions of the \mbox{SDSS-III} Collaboration including the University of Arizona, the Brazilian Participation Group, Brookhaven National Laboratory, Carnegie Mellon University, University of Florida, the French Participation Group, the German Participation Group, Harvard University, the Instituto de Astrofisica de Canarias, the Michigan State/Notre Dame/JINA Participation Group, Johns Hopkins University, Lawrence Berkeley National Laboratory, Max Planck Institute for Astrophysics, Max Planck Institute for Extraterrestrial Physics, New Mexico State University, New York University, Ohio State University, Pennsylvania State University, University of Portsmouth, Princeton University, the Spanish Participation Group, University of Tokyo, University of Utah, Vanderbilt University, University of Virginia, University of Washington, and Yale University.

\bibliographystyle{apj}
\bibliography{refs}

\end{document}